\documentclass[sigconf]{acmart}   

\usepackage{subcaption}
\usepackage{xspace}
\usepackage{multirow}
\usepackage{dblfloatfix} 

\AtBeginDocument{%
  \providecommand\BibTeX{{%
    \normalfont B\kern-0.5em{\scshape i\kern-0.25em b}\kern-0.8em\TeX}}}

\copyrightyear{2024}
\acmYear{2024}
\setcopyright{rightsretained}
\acmConference[CHI '24]{Proceedings of the CHI Conference on Human Factors in Computing Systems}{May 11--16, 2024}{Honolulu, HI, USA}
\acmBooktitle{Proceedings of the CHI Conference on Human Factors in Computing Systems (CHI '24), May 11--16, 2024, Honolulu, HI, USA}
\acmDOI{10.1145/3613904.3642217}
\acmISBN{979-8-4007-0330-0/24/05}

\definecolor{darkgreen}{rgb}{0,0.5,0}
\definecolor{orange}{rgb}{1,0.5,0}
\definecolor{teal}{rgb}{0,0.5,0.5}
\definecolor{darkpurple}{rgb}{0.5, 0, 0.5}
\definecolor{olive}{rgb}{0.6,0.6,0}

\newcommand {\michael}[1]{ }
\newcommand {\piyawat}[1]{ }
\newcommand {\canl}[1]{ }
\newcommand {\bjoern}[1]{ }
\newcommand {\jeremy}[1]{ }
\newcommand {\jd}[1]{ }
\newcommand {\matthew}[1]{ }
\newcommand {\su}[1]{ }

\newcommand{\tool}[1][]{\textsc{Rambler}#1\xspace}

\newcommand{\edits}[1]{#1}
\newcommand{\newedit}[1]{#1}
\newcommand{\deletes}[1]{}

\begin{document}

\graphicspath{{./figures}}

\title{\tool: Supporting Writing With Speech via LLM-Assisted Macro Revision}
\title{\tool: Supporting Interactive Dictation and Revision with LLM-backed Semantic Manipulations}
\title{\tool: Supporting Writing With Speech via LLM-Assisted Semantic Manipulations}
\title{\tool: Supporting Writing With Speech via a Gist-based Interface}
\title{\tool: Supporting Writing With Speech via LLM-Assisted GUI for Revision}
\title{\tool: Writing with the Gist of Speech via a LLM-augmented GUI }
\title{\tool: Speech-based Long-form Writing via a LLM-Augmented GUI}
\title{\tool: Supporting Writing With Speech via LLM-Assisted Gist Manipulation}



\author{Susan Lin}
\affiliation{%
  \institution{UC Berkeley}
  \city{Berkeley, CA}
  \country{USA}
}
\author{Jeremy Warner}
\affiliation{%
  \institution{UC Berkeley}
  \city{Berkeley, CA}
  \country{USA}
}
\author{J.D. Zamfirescu-Pereira}
\affiliation{%
  \institution{UC Berkeley}
  \city{Berkeley, CA}
  \country{USA}
}
\author{Matthew G. Lee}
\affiliation{%
  \institution{UC Berkeley}
  \city{Berkeley, CA}
  \country{USA}
}
\author{Sauhard Jain}
\affiliation{%
  \institution{UC Berkeley}
  \city{Berkeley, CA}
  \country{USA}
}
\author{Michael Xuelin Huang}
\affiliation{%
  \institution{Google}
  \city{Mountain View, CA}
  \country{USA}
}
\author{Piyawat Lertvittayakumjorn}
\affiliation{%
  \institution{Google}
  \city{Mountain View, CA}
  \country{USA}
}
\author{Shanqing Cai}
\affiliation{%
  \institution{Google}
  \city{Mountain View, CA}
  \country{USA}
}
\author{Shumin Zhai}
\affiliation{%
  \institution{Google}
  \city{Mountain View, CA}
  \country{USA}
}
\author{Björn Hartmann}
\affiliation{%
  \institution{UC Berkeley}
  \city{Berkeley, CA}
  \country{USA}
}
\author{Can Liu}
\affiliation{%
  \institution{School of Creative Media, City University of Hong Kong}
  \city{Hong Kong}
  \country{China}
}
\authornote{Corresponding author}

\renewcommand{\shortauthors}{Lin, et al.}

\begin{abstract}
Dictation enables efficient text input on mobile devices. However, writing with speech can produce disfluent, wordy, and incoherent text and thus requires heavy post-processing. This paper presents \tool, an LLM-powered graphical user interface that supports gist-level manipulation of dictated text with two main sets of functions: gist extraction and macro revision. Gist extraction generates keywords and summaries as anchors to support the review and interaction with spoken text. LLM-assisted macro revisions allow users to respeak, split, merge, and transform dictated text without specifying precise editing locations. Together they pave the way for interactive dictation and revision that help close gaps between spontaneously spoken words and well-structured writing. In a comparative study with 12 participants performing verbal composition tasks, \tool outperformed the baseline of a speech-to-text editor + ChatGPT, as it better facilitates iterative revisions with enhanced user control over the content while supporting surprisingly diverse user strategies.
\end{abstract}

\ccsdesc[500]{Human-centered computing~Interactive systems and tools}
\ccsdesc[500]{Human-centered computing~Interaction paradigms}
\ccsdesc[300]{Human-centered computing~Interaction techniques}
\ccsdesc[100]{Human-centered computing~Natural language interfaces}

\keywords{dictation, speech, speech-to-text, STT, text composition, writing, LLM, AI}

\begin{teaserfigure}
\centering
\includegraphics[width=.75\textwidth]{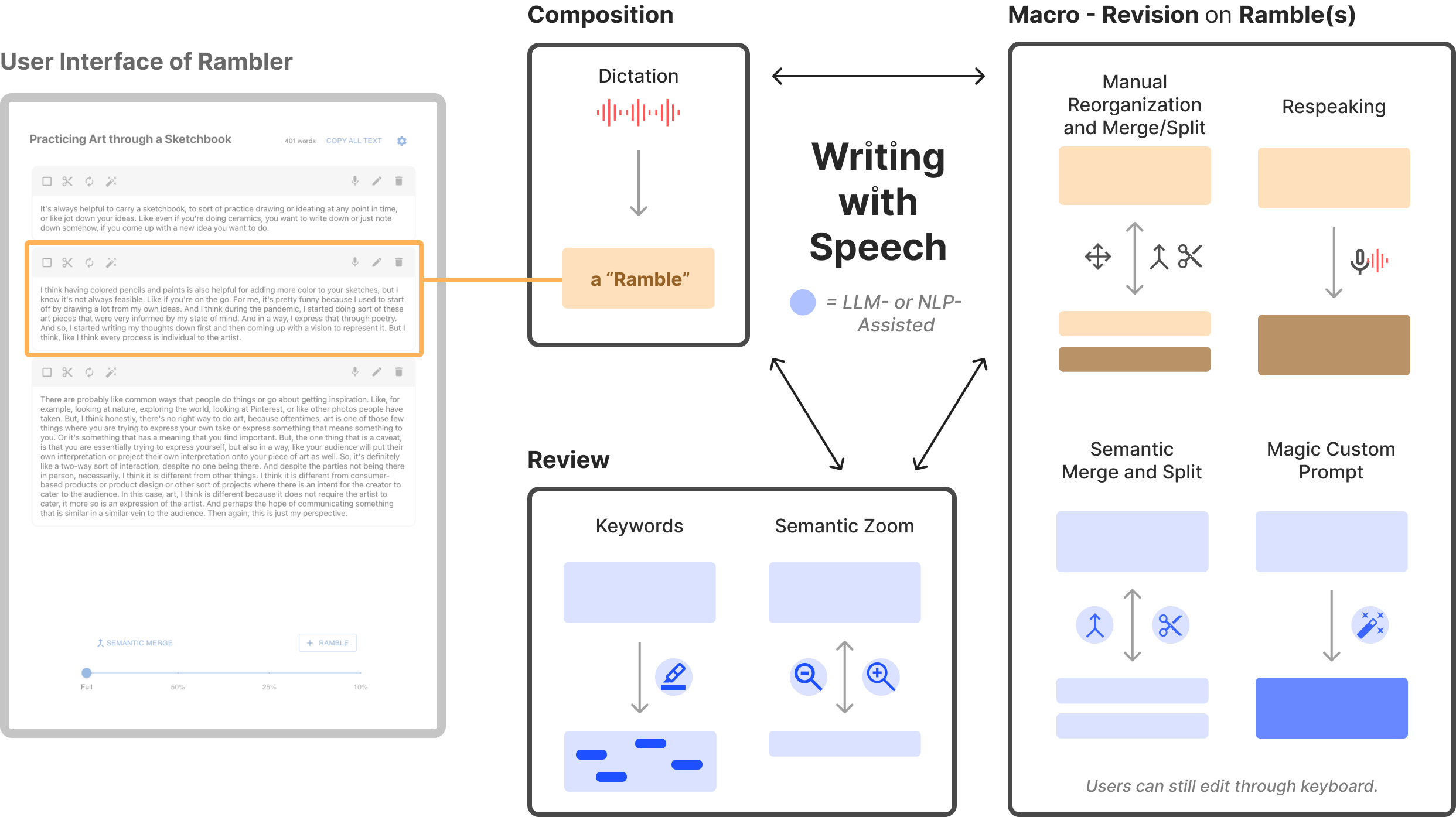}
  \caption{Supporting long-form text composition via speech with a new dictation interface that consists of three components: 1) dictating trains of thought into Ramble boxes, 2) reviewing the spoken text through automatically generated keywords and summaries, and 3) performing manual or LLM-assisted macro-revisions operated at conceptual levels. } 
  \label{fig:teaser}
\end{teaserfigure}

\maketitle


\section{Introduction}

Speech, regarded as a natural input modality, has attracted much attention following the breakthroughs of automatic speech recognition (ASR) and natural language processing (NLP). Speech transcription software products like Otter.ai, Google Voice Typing, and Dragon are gaining popularity. Dictation is now an integral keyboard function on most smartphones and tablets. 
Indeed, dictation can be three to four times faster than typing for text input under the best circumstances~\cite{10.1145/3161187}. This makes it particularly advantageous for inputting long text, such as writing memos or drafting ideas. Mobile devices benefit even more, given the challenges of using a keyboard and cursor on a touchscreen~\cite{vogel2007shift}. 

Current interfaces for dictating long text largely mirror traditional text editors with an added microphone. However, \emph{writing with speech can be more than just dictation.} 
Previous research \cite{10.1145/3571884.3597134} showed that speaking differs significantly from typing in terms of both production and memory retention. Transcribed spoken text can be much more verbose, erroneous, and disorganized, leading to high-effort editing \cite{10.1145/302979.303160}. The verbatim memory of spoken content is also weaker, leading to a likely higher effort in reading the transcripts. Therefore, treating speaking simply as a faster ``typing'' method can be problematic, especially for long texts. Similarly, researchers of voice user interfaces (VUI) have long argued against directly transplanting GUI guidelines to VUI design~\cite{8794686}. All these call for new affordances and interface solutions for writing with speech. 
Further, writing is a highly iterative activity and thought development process. Typically, digital writing requires people's full attention on a screen while sitting in front of a computer with a mouse and keyboard, but this requires time and an environment, which creates a high barrier to starting. While speech has the potential to be used for faster and lower-barrier writing, its adoption remains limited and the effort to edit resulting transcripts is large. 

This work sets out to tackle that challenge and answer the question: \emph{Can we identify and build stepping stones towards closing the gap between speaking and writing, by designing a dictation interface that addresses the challenges inherent in speech?} We build on existing solutions in recent publications that use NLP and large language models (LLMs) to generate summaries to assist writing \cite{Dang22} and review spoken dialogue \cite{10.1145/3472749.3474771,li2023improving}. These interfaces extract the gist of text to support visualization and/or interaction with that text, a paradigm we call \emph{gist-based interfaces}. They move beyond automatic text summarization to semantic manipulation, with the interaction centered around the ``gist'' of each piece of text, rather than the sequences of characters in each word of that text. While such concepts have been explored in traditional writing \cite{Dang22}, we operationalize these ideas in the context of interacting with spoken text, which is perhaps an even more compelling use given the literature that shows spoken content more often remains in memory in the form of gists, rather than verbatim \cite{10.1145/3571884.3597134}.  

We introduce \tool, an LLM-powered graphical user interface (GUI) that implements a gist-based interface tailored to dictated text (see Figure \ref{fig:teaser}). Users can dictate their extemporaneous thoughts and place each segment of recording (i.e., a fragment of an idea) into a \textit{Ramble}.  \tool provides \textit{Semantic Zoom}~\cite{perlinZoom} to visualize multiple summarization levels in the form of gists for each Ramble, which users can review on demand. 
Users then focus on iterating on higher-level ideas by manipulating the Rambles, through respeaking, splitting, merging and reorganizing them, activities that we collectively call \textit{macro revision}. 
Behind the scenes, \tool uses an LLM to automatically clean up any disfluencies and punctuation errors from the raw transcript, 
as well as to complete broken sentences and to smooth transitions, activities that we call \textit{micro revision}. 
In addition, \tool extracts keywords from text to aid visual skimming across Rambles and Semantic Zoom levels. Individual keywords can also be activated and deactivated by the user to indicate importance; these keywords then serve as parameters to customize summary (gist) generation and other LLM-based word processing.

\tool aims to aid the review and revision of spoken text and ease interactions with text on mobile devices. To evaluate \tool's effectiveness, we conducted a comparative study with 12 participants using a long-form verbal text composition task. The baseline is a standard speech-to-text editor interface with ChatGPT provided on the side. While participants demonstrated varied strategies and feature usage, most of them expressed a preference of \tool over the baseline and acceptance of using it for their own tasks. Their feedback shows \tool provides better support for developing and reviewing spoken content, supporting iteration, and improving user control over content revision.

Our work makes the following contributions:
First, \tool: a novel tool that implements the concept of a gist-based interface to support writing with speech on mobile devices. We motivate our design goals and provide implementation details for \tool.    
Second, we contribute empirical findings from a user study revealing the potential of \tool in helping close the gap between rambling and writing. Our study also reveals several advantages of our specialized LLM-backed GUI over a one-size-fits-all dialogue-based chatbot UI for LLMs; we thus also offer insights, based on our findings, for designing and building LLM-supported \edits{direct} manipulation interfaces. 

\section{Related Work}
To provide more context for our contributions, we discuss three areas of highly related work: 
voice dictation and editing, supporting writers with LLMs, and semantic manipulations on text.

\subsection{Voice Dictation and Editing}

Although speaking is faster than typing or writing, it can be time-consuming to correct errors caused by speech artifacts, such as disfluencies and repetitions. Modern speech and NLP research has made much progress in cleaning disfluency and speech recognition errors \cite{liao2023improving,lou2020end,tanaka2018neural}, mostly for post-processing. Researchers have also proposed multimodal interaction methods to reduce editing effort, especially on mobile devices. For example, EyeSayCorrect uses gaze to infer users' editing intentions and allows re-dictation to edit the phrase near eye fixations~\cite{10.1145/3490099.3511103}. Similar ideas were applied to other modalities, such as touch~\cite{zhao2021voice}. 

Other work has explored speech-based editing through automatic removal of colloquial or filler words~\cite{arnold2021generative} as well as voice editing with and without explicit command keywords~\cite{10.1145/3390889}. Ghosh et al.~\cite{10.1145/3390889} found that using explicit command words (e.g., "replace" and "delete") were suitable for one-word edits, and re-dictation without command words was suitable for complex edits. To leverage the advantages of both cases, "Just speak it" supports commanding without explicit command words~\cite{10.1145/3472749.3474795}, which are instead induced from existing context and the edit command. Li et al.~\cite{li2023interactive} further explored auto segmentation and classification to support the seamless transition between voice dictation and editing with natural language. While this line of work lays the groundwork for natural and effortless voice-based text editing, it primarily focuses on editing one or a few words. Generalizing to sentence-level macro revision remains largely unexplored.

\subsection{Writing with LLMs}

A number of studies have demonstrated that text generation by LLMs can be beneficial for writing assistance. For example, Yuan et al.~\cite{Coenen2021WordcraftAH} developed Wordcraft, an editor that allows users to collaborate with an LLM to write stories. Users can replace selected text with LLM suggestions or prompt the model for text generation. 
Yang et al.~\cite{Yang2022AIAA} investigated an LLM-powered writing tool to shorten, edit, summarize, and generate text. 
Gero et al.~\cite{10.1145/3532106.3533533} customized the decoding method of an LLM to encourage diverse outputs specifically for scientific writings.

While LLMs have made significant progress in the field of writing, there is still room for improvement. LLMs may struggle to retain the writing style and align with the writer's goal. Ippolito et al.~\cite{Ippolito2022CreativeWW} found that LLMs can struggle to preserve the style and authorial voice of experienced writers. This is supported by Biermann et al.'s~\cite{Biermann2022FromTT} study, which found that authors prefer text generation tools that respect their styles and strategies. Rahman et al.~\cite{Rahman2023ChatGPTAA} studied academic writing with ChatGPT. Their findings revealed that ChatGPT can be useful to generate initial ideas, but falls short in tasks that demand critical thinking, such as literature synthesis, citations, problem statements, research gaps, and data analysis. Taken together, LLMs are a promising tool for writing, but they alone are not yet perfect. 

\subsection{Semantic Manipulations on Text}

In addition to text generation discussed in the previous section, LLMs can also support macro-level structural revision. However, as Strobl et al.~\cite{strobl2019digital} recently surveyed, writing tools that support grammar and word-level editing are common, but tools for macro-level structural revision are relatively rare. Arnold et al.~\cite{arnold2021generative} highlighted the opportunities to return semantic control to the writers. One of their proposals is to support within-sentence structural manipulation, where users can drag and drop words to rephrase the sentence.
Alongside that, semantic zooming~\cite{perlinZoom} is another promising feature for collaborative human-AI interfaces. It allows users to explore and navigate complicated data (e.g. long texts and nested tables) by zooming in and out to check details of different levels. 
This can be particularly useful for interacting with LLMs in a conversational paradigm, as it reduces the cognitive and physical load of scrolling back and forth to linearly search for information from a lengthy conversation.
Several studies have demonstrated that semantic zooming is an effective way to help users navigate and digest complex content. Sensescape~\cite{suh2023sensecape} and Graphologue~\cite{jiang2023graphologue} introduce zoomable visual representations of different concepts involved in a conversation with LLMs. Li et al.~\cite{10.1145/3472749.3474771, li2023improving} developed a system that generates hierarchical summaries of spoken dialog. 
As the closest work to ours, Beyond Text Generation~\cite{Dang22} is a writing tool that provides on-the-fly paragraph summarization along with the original text. This work also provides interaction with the summaries of paragraphs, such as reorganization via drag and drop, which manipulates the original text in parallel. Building on this work, we leverage the benefit of summaries as one way of extracting gists. While its focus is on keyboard typing and ours is on speech input, our interface also provides keywords extraction and interaction as well as other LLM-assisted operations in order to support macro revision.

\section{Rambler: A Gist-based Interface for Writing With Speech}

\subsection{Design Rationale and Goals}

As explained in the introduction, we aim to design a gist-based dictation interface to support writing with speech, by helping users overcome the challenges of speech production and shortcomings of memory. The challenges and shortcomings identified in previous empirical and theoretical work~\cite{10.1145/3555758, 10.1145/3571884.3597134} can be summarized as follows: 1) speech is spontaneous as it is produced in real-time, so while the gists may be planned, the wording is not; 
2) looking at a live transcript while speaking can be distracting, so people tend to have less visual engagement speaking than they would typing, which leads to less micro-revision when speaking;
3) remembering spoken content verbatim is a challenge memory task; combined with reduced visual engagement, these challenges may also make reviewing content harder. 

\begin{figure*}[t]
    \centering
    \includegraphics[width=0.9\textwidth]{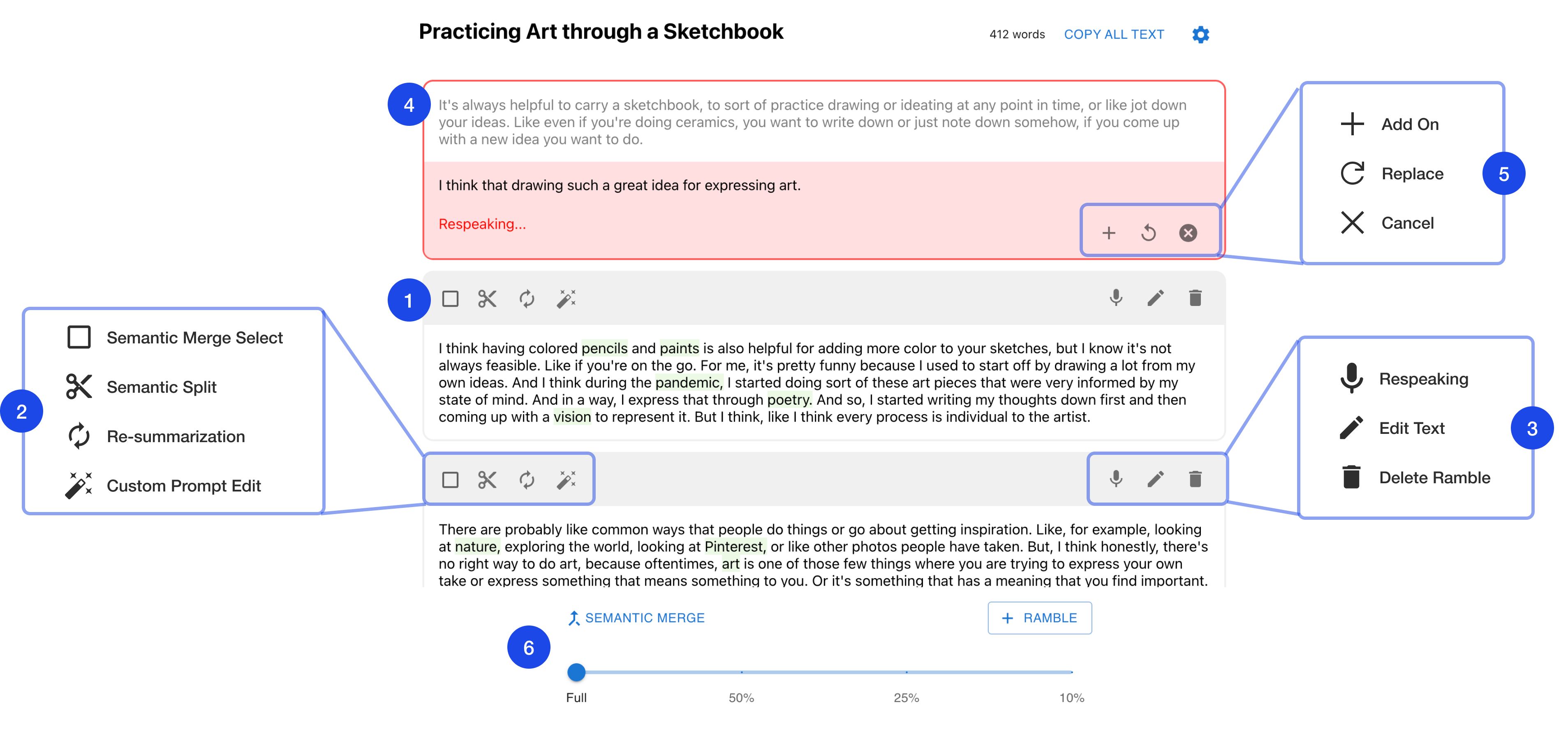}
    \caption{A labeled screenshot of the \tool UI. (1) Ramble in default state, with revision functions accessible through buttons on (2) and (3). (4) Ramble in re-speaking mode, where voice input is transcribed so that it can be appended to current text, replace the current text, or to be discarded using the buttons on (5). 
    Fixed at the bottom of the UI is (6), with the Semantic Merge button, New Ramble button, and Semantic Zoom slider.}
    \label{fig:ui-screenshot}
\end{figure*}
\subsubsection{Rambles as the unit of interaction} \label{sec:rambles-interaction}

With these characteristics, a text editor that transcribes everything linearly will also include all the disfluencies, repetition, and disorganization of raw speech. Although advanced NLP techniques can help correct some of these issues, the user still needs to direct the composition's meaning and organization. Successful LLM interfaces (e.g., ChatGPT,\footnote{https://chat.openai.com/} Claude\footnote{https://claude.ai/}) have demonstrated the ability to refine unorganized text into high-quality writing. However, such general tools not only rely on users' ability to prompt the LLM, which has been observed to be challenging \cite{zamfirescu2023prompt, zamfirescu2023herding}, but also make it harder for users to precisely control where and how their text is edited. 

Building on previous work that shows the benefits of using NLP or LLMs in direct manipulation interfaces~\cite{Dang22,jiang2023graphologue}, we investigate how to leverage recent LLM and NLP techniques to inject intelligent word processing into the GUI for our specific task of writing with speech, with the goal of preserving user control over the composition. With these considerations, we aim to improve the text editor interface for better interaction with spoken text by establishing the following design goals:

\begin{itemize}
    \item \textit{Support non-linear composition.} The tool should allow users to \textit{ramble} spontaneously while providing them with structures to support iterative organizing and editing their spoken text. 
    
    \item \textit{Support iterative drafting.} Leverage the fast production of speech while recognizing that the first attempt may not be well articulated as written text. Our interface should encourage and support users easily iterating on drafts by respeaking large chunks of the text before diving into micro-editing. 

    \item \textit{Aid the review and navigation of spoken text.} The interface should help reduce user effort in reading and comprehending the transcript, which could be verbose, repetitive, and contain errors or disfluencies.

    \item \textit{Leverage LLMs while preserving user control.} The interface should leverage the capacity of LLMs in cleaning up, smoothing, and transforming text while allowing users to control where and how changes will be applied. 
    
    \item \textit{Optimize interactions for mobile devices.} Although speech-tor-text input can be used on any device, mobile devices particularly benefit, given the known challenges of interacting with text on \edits{these devices without keyboards}. For our evaluation, we target a tablet interface for designing our current interface, \edits{as it is a commonly used mobile device suitable for long-form writing. } 

\end{itemize}

In the next subsection, we will explain how we designed the interface of our tool to achieve these design goals.

\begin{figure*}
    \centering
    \includegraphics[width=0.98\textwidth]{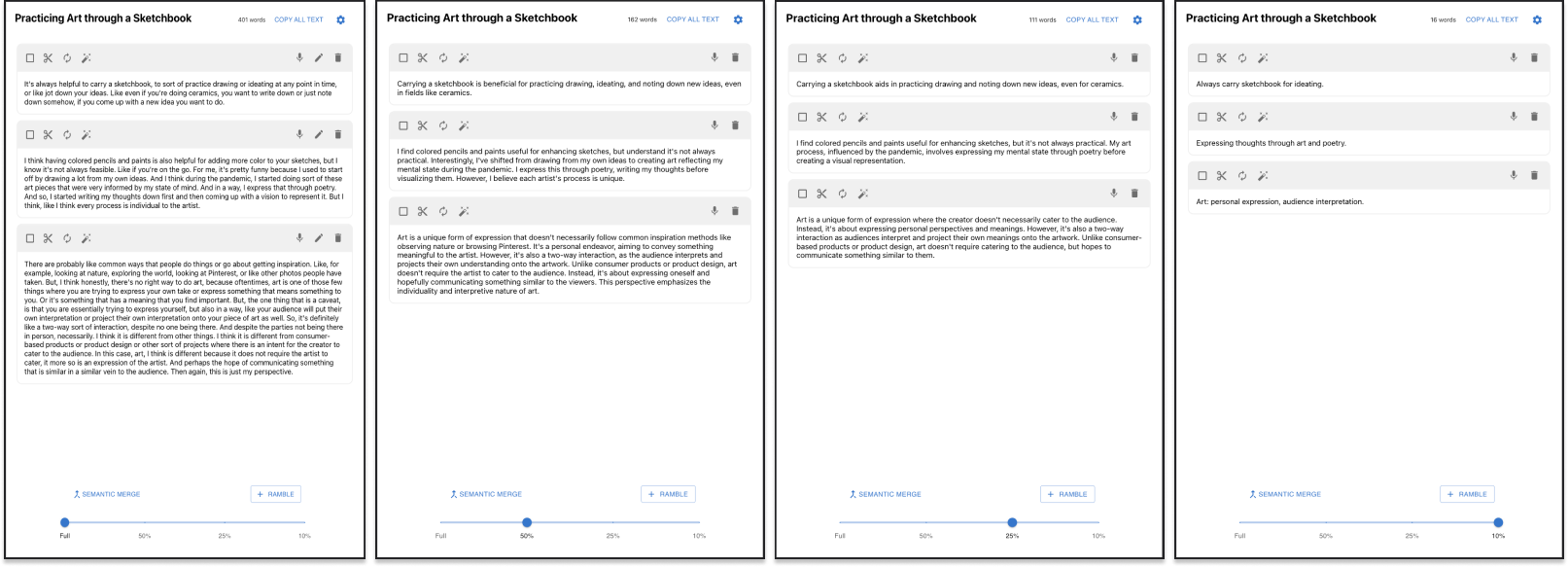}
    \caption{From left to right, example text in three Rambles is presented at all four Semantic Zoom levels: full transcript, 50\% length, 25\% length, and 10\% length.}
    \label{fig:semanticzoom}
\end{figure*}

\begin{figure*}
\centering
  \begin{subfigure}{0.49\linewidth}
    \centering
    \includegraphics[width=\linewidth]{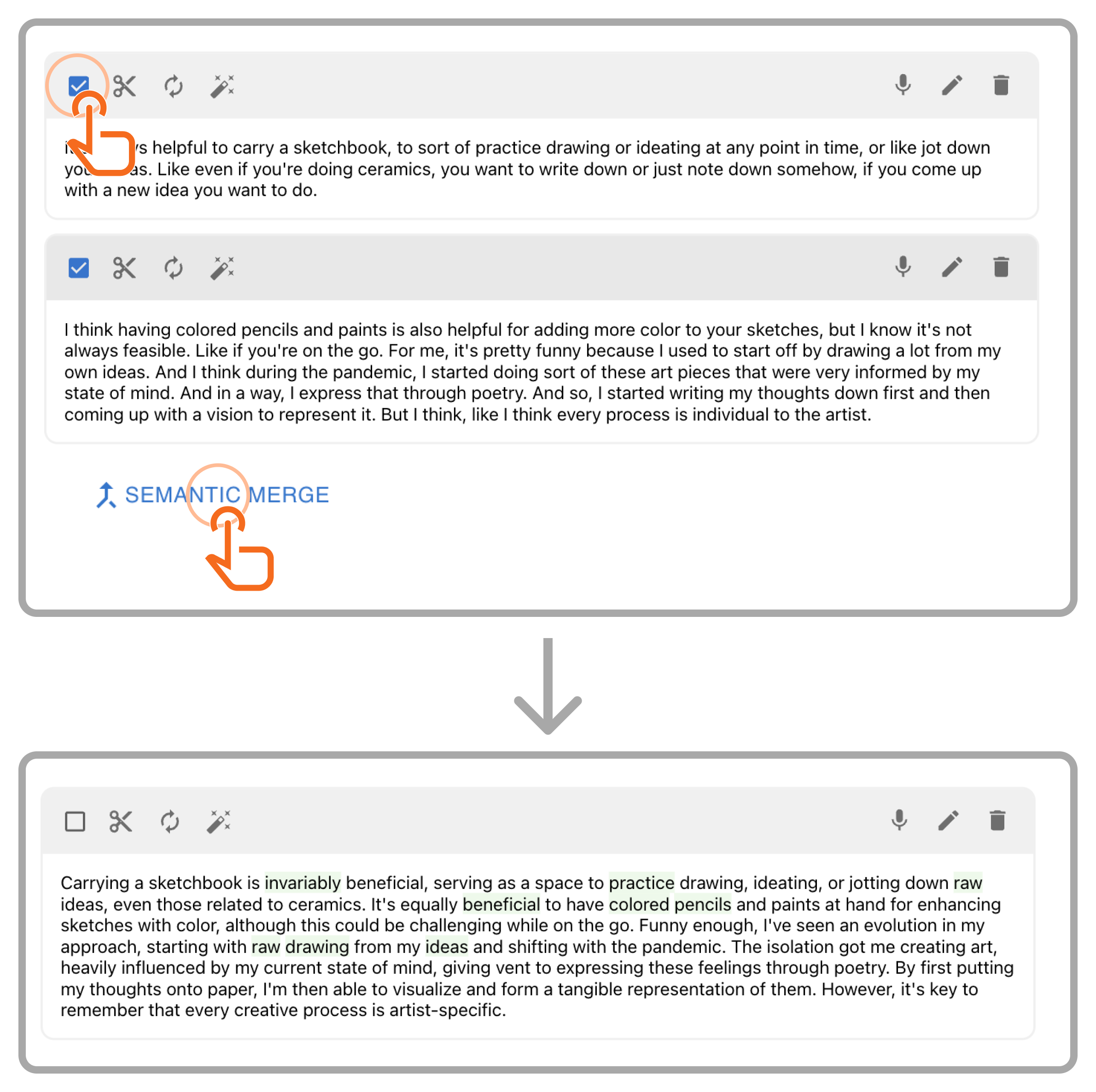}
    \caption{}
  \end{subfigure}%
  \hfill
  \begin{subfigure}{0.49\linewidth}
    \centering
    \includegraphics[width=\linewidth]{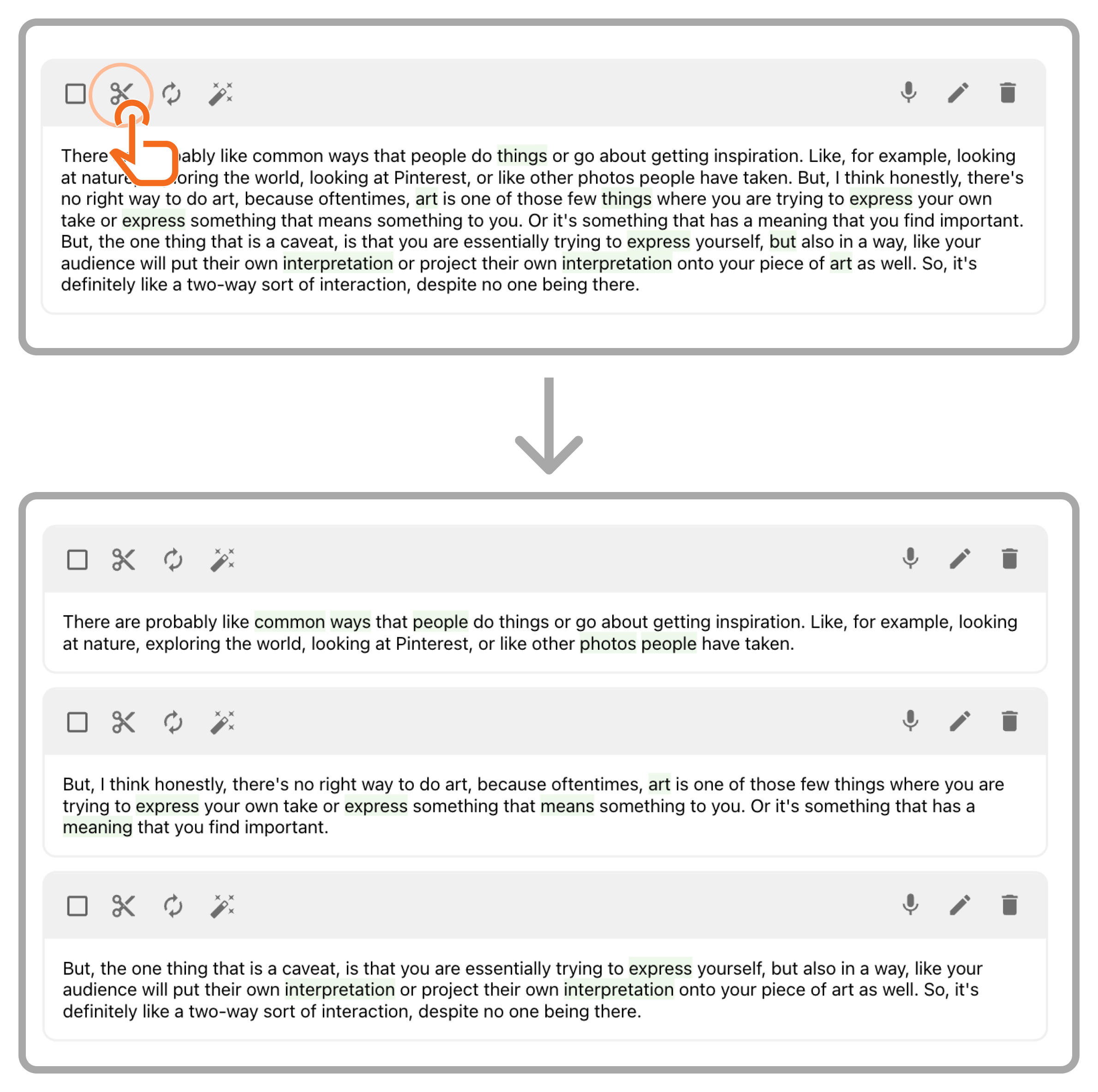}
    \caption{}
  \end{subfigure}%
  \caption{\edits{Example Rambles before and after (a) Semantic Merge and (b) Semantic Split. In (a), the user select all the Rambles to include, then press the Semantic Merge button (shown in Figure \ref{fig:ui-screenshot}). In (b), the user taps the scissors icon in a Ramble to ask LLM to split it based on content. }}
    \label{fig:semantic_merge_split}
\end{figure*}

\subsection{\tool Interface Design}

\tool is built around the creation and manipulation of \textit{Rambles}, illustrated in Figure \ref{fig:teaser} and \ref{fig:ui-screenshot}. Each \textit{Ramble} is a container (visually separated from other Rambles) that holds its content and multiple buttons for users to interact with.
Our intention is for the content of a Ramble to roughly correspond to a single train of thought. 
Without the pressure of producing a decent piece of text in one shot, users may either speak everything into one Ramble and split it later, or start a new Ramble when they feel what they will say next is about a different point or subject matter. \tool should support various writing strategies, including those starting both with or without an outline in mind. Here, we explain the design of \tool and its main features supporting macro revision and text reviewing, respectively.

\subsubsection{Supporting text reviewing with clean transcript and gist extraction} \label{sec:text-reviewing}
\tool is designed to help users review dictated text more efficiently with three key features
(transcript cleaning, semantic zooming, and highlighting keywords) that leverage NLP and LLMs.

First, \tool uses an LLM to automatically clean Ramble transcripts when they are added or edited, in order to minimize small speech-to-text (STT) typos (like messy punctuation or misheard words) that impact reading. To avoid distraction, our interface only shows raw speech for real-time feedback when the user is actively dictating; if the user stops dictating, the interface immediately cleans the spoken text, and only displays the clean transcript.

Second, we implement Semantic Zoom using a slider that controls the degree of summarization shown across all the Rambles in a composition. As shown in Figures~\ref{fig:ui-screenshot} (component 6) and~\ref{fig:semanticzoom}, users can move the slider at the bottom of the UI to view text summaries at different levels, 
specifically, at 50\%, 25\%, and 10\% of the original length.\footnote{\tool's prompts for these summary levels were tuned to consistently achieve these levels, but are not explicit about these percentages; see appendix~\ref{adx:summarize}.}
We designed this feature to help visualize the core ideas of lengthy text and to allow users to quickly gain a sense of how their ideas should be re-organized and iterated upon.

Last, to aid visual skimming across Rambles, \tool automatically highlights keywords after a Ramble is added or edited (see Figure~\ref{fig:keyword-magic-prompt}-top).
Users can also tap any word in a Ramble to manually deselect or select the word as a keyword. 
Keywords are not only a visual aid for continuity across zoom levels; users can also tap the \textit{regenerate} button on a Ramble to re-summarize their transcript text based on their modified keyword selection. This helps address an issue we discovered where summaries generated by LLMs may not be aligned with the users' main ideas in a Ramble. 

\subsubsection{Supporting macro revision with LLM assistance} \label{sec:macro-revision}

We distinguish between conceptual-level edits, which we call macro-revisions, from fine-grained edits, which we call micro-revisions. Macro-revisions operate on ideas, and in our prototypes happen at a Ramble level – consider functionalities like reorganizing Rambles, and merging or splitting Rambles. Micro-revisions, in contrast, operate on a much smaller scale and, in our prototypes, happen primarily through keyboard editing – consider fixing a single typo or some word choice in a sentence. We consider macro-revisions and micro-revisions complementary, thus our interface also provides keyboard editing, activated by clicking the \textit{edit} button in a Ramble.

We designed five macro-revision operations: Ramble respeaking, Ramble re-ordering through drag and drop, Ramble merging (concatenation of text) by dragging one Ramble onto another, Ramble splitting (by pressing return during keyboard editing), and Ramble transformation. 
Although macro-revisions can also be done in a normal text editor, we provide an interface with a stronger affordance for doing so, with the goal of encouraging and supporting users to apply these operations on spoken text. 
We carefully designed a re-speaking mechanism and interface (Figure~\ref{fig:ui-screenshot}, zone 4), which is activated when the user re-activates the mic of an existing Ramble. The user talks further and then chooses whether to append the new text to the Ramble, replace the old text in the Ramble with it, or discard the new text (Figure~\ref{fig:ui-screenshot}, zone 4). During the respeaking process, the original text remains above in grey color as a reminder. We do so to aid the user's memory, as in our design process we learned that without the original text, it is hard to remember what to say.
 
\begin{figure}
    \centering
    \includegraphics[width=\columnwidth]{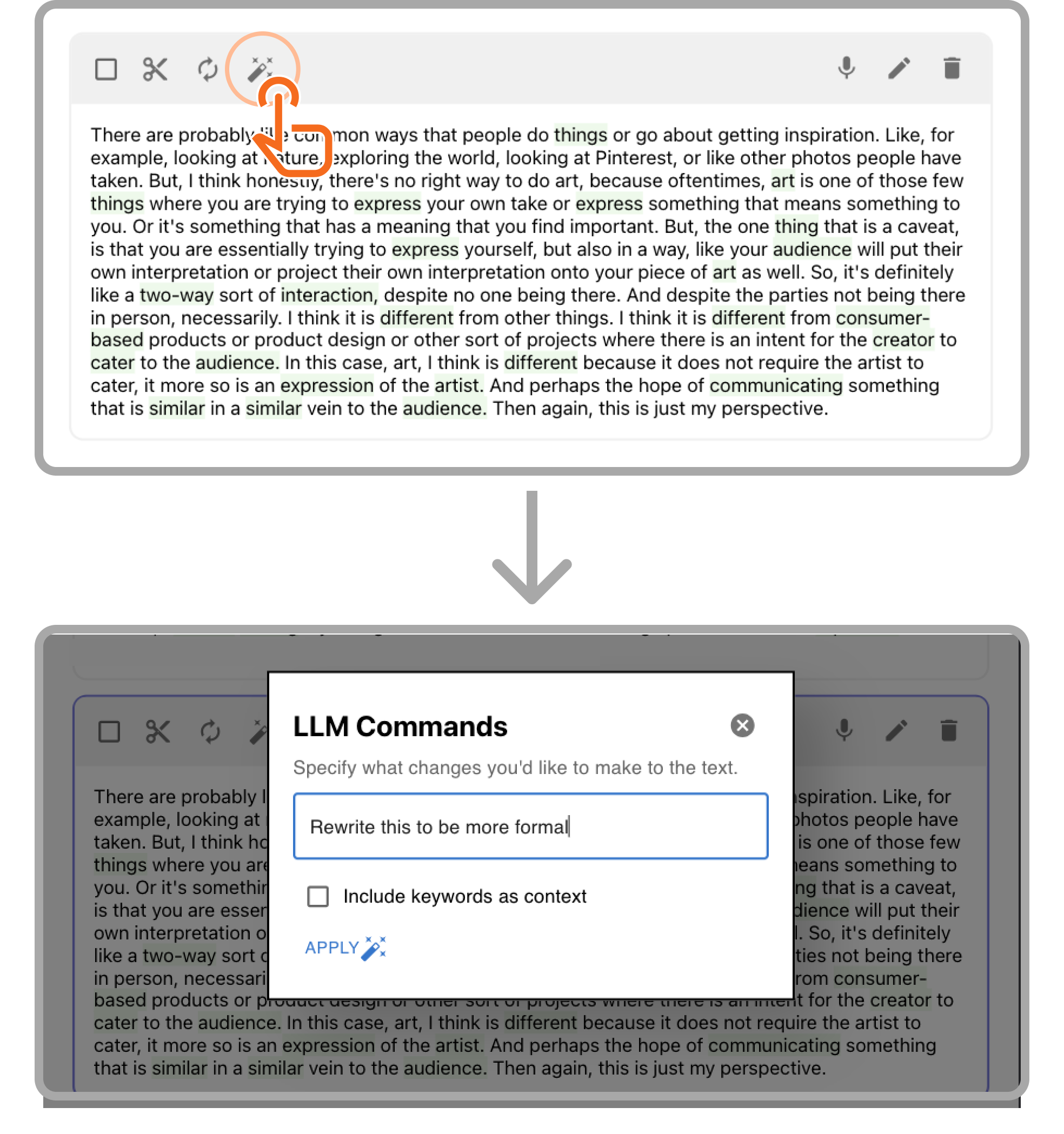}
    \caption{Example transforming a ramble with Magic Custom Prompt: (top) keywords highlighted in light green; (bottom) custom prompt window triggered by the magic wand button on a Ramble. The user inputs an example prompt, and optionally ticks the checkbox for including keywords as context.}
    \label{fig:keyword-magic-prompt}
\end{figure}

While all of the macro revisions can be performed manually, we provide an LLM-assisted version for three of the operations -- Ramble merge, split, and transformation -- and named them \emph{Semantic Merge, Semantic Split,} and \emph{Magic Custom Prompt}.
Semantic merging utilizes the LLM to intelligently merge the content \edits{of} multiple selected Rambles into the content of one Ramble by clicking a \textit{Semantic Merge} button (see Figure \ref{fig:ui-screenshot}.6). Semantic Split, represented by a \textit{scissors} button, is placed on individual Rambles, which prompts the LLM to divide one particular Ramble into N Rambles based on its content. These two mechanisms are visualized in Figure \ref{fig:semantic_merge_split} with example text. The Magic Custom Prompt feature is triggered by a \textit{magic wand} button on a Ramble. It opens up the possibility for expert LLM users to define any custom transformation on Ramble content by directly inputting an LLM prompt (see Figure \ref{fig:keyword-magic-prompt}).  


\subsection{Example User Workflow}

To better illustrate the functionality of \tool, consider the following example of a simple user workflow: Alex, an artist, is drafting a post for his personal blog and wants to describe how he finds creative inspiration. He has many ideas on the topic, but isn't sure the best way to write it: hence, he reasons that he should simply talk it out. Opening up \tool, Alex presses the \textit{+ Ramble} button to begin recording.
After finishing his thoughts and stopping the recording, \tool automatically cleans his dictated text into the first Ramble. Alex senses that he wandered somewhat from his initial topic; since the resulting text is fairly long, he slides the Semantic Zoom slider to see a summarized version of his text. Then he takes a look at the highlighted keywords in all the zoom levels to get a better sense of what he talked about. After glancing back at the transcript, Alex decides to use Semantic Split to revise the various ideas in his Ramble separately.

\tool segments his cleaned spoken text into multiple rambles along three main ideas, 1) ``finding personal inspiration,'' 2) ``interaction between self expression and audience interpretation,'' and 3) ``difference between art and design.'' With a clearer understanding of the ideas he's just verbalized, Alex begins to alternate between composing new Rambles, and revising with the provided functionalities (e.g., respeaking, reordering, merging, and/or splitting Rambles) -- until he feels like each Ramble roughly resembles a somewhat concrete idea. Once he feels that he has added all his thoughts and developed them into a cohesive flow of ideas, he uses the built-in onscreen keyboard to fix word choices here and there to his liking, and then exports his composition to share with the world.  

\begin{figure}
    \centering
    \includegraphics[width=\columnwidth]{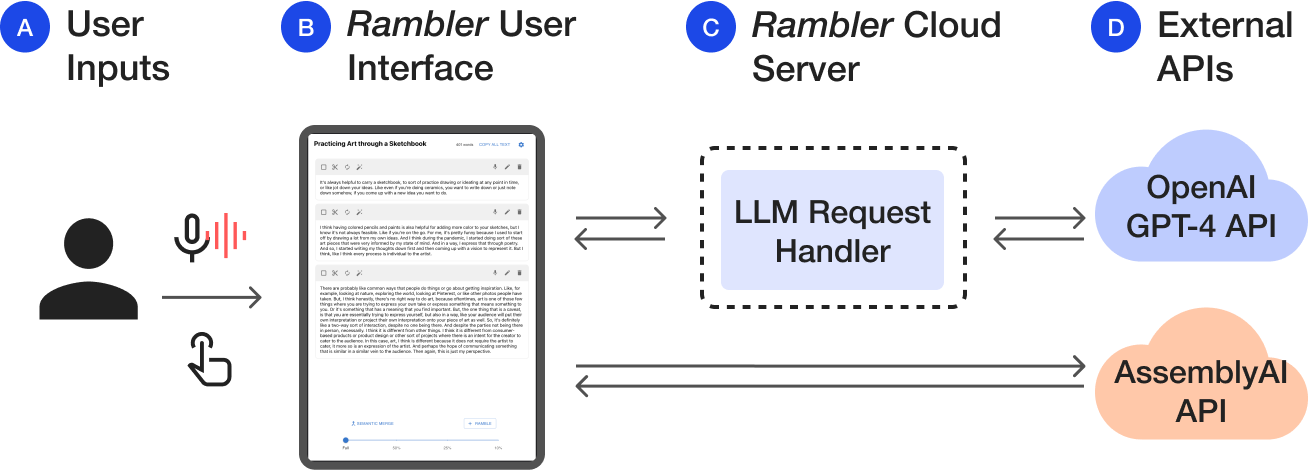}
    \caption{\tool System Architecture Diagram consisting of a web-based frontend, a cloud server mediating between user requests and the OpenAI API for LLM functionality. The AssemblyAI API is used for real-time speech transcription.}
    \label{fig:data-flow}
\end{figure}

\subsection{Prototype Implementation} \label{sec:temp-impl}

\tool is implemented as a React.js web application with a node.js backend and \edits{is} hosted in the cloud. LLM-based text manipulation functionality is enabled using OpenAI's GPT-4 API, which provides high quality output for our desired LLM functionality. We considered other models such as GPT-3.5 Turbo for response time reasons; their output on LLM interactions (most notably, Semantic Split) was less precise. \tool's voice transcription uses AssemblyAI’s real time Speech-to-Text API. We made this engineering choice because AssemblyAI's API allowed us to transcribe text indefinitely in a single instance. Automatic keyword selection is implemented via the rapid automatic keyword extraction (RAKE) method \cite{rose2010rake}. Compared to an LLM based approach for keyword selection akin the method used in Graphologue \cite{jiang2023graphologue}, the RAKE method was orders of magnitude faster ($\sim$4s vs nearly instant). Our overall system architecture is illustrated with Figure \ref{fig:data-flow}. Our project source code is available at \href{https://github.com/BerkeleyHCI/rambler}{https://github.com/BerkeleyHCI/rambler}.

We used GPT-4 to power the four key LLM-based gist-related features: text summarization for Semantic Zoom (including transcription cleaning), merging multiple Rambles, splitting a single Ramble into multiple, and open-ended editing using customized LLM prompts. We created pre-defined prompts for zoom, merging, and splitting; the open-ended editing prompts are entirely user defined. One-shot prompting was employed for the transcription cleaning prompt and zero-shot was used for the rest. Keywords are appended to prompts when marked for inclusion through selection. All prompts are attached in Appendix~\ref{app_prompts} for reference.

\section{Evaluation} 

To evaluate the effectiveness of \tool for supporting writing with speech, we designed a \textit{within-subjects} study to compare \tool with a state-of-the-art solution for dictation and word processing. We chose, as our \textbf{baseline}, \emph{raw dictation into a text editor with ChatGPT provided on the side}. Participants completed two long-form writing tasks through dictation, one using \tool's full suite of available functionality, and the other using the baseline. 
Through this comparison, we seek answers to the following questions:

\begin{enumerate}
    \item Is writing with speech better supported by \tool than the baseline? If so, how? 
    \item How effectively does our approach of embedding LLMs in a GUI support the review and revision of spoken text? 
 \end{enumerate}

\subsection{Baseline Interface}

As mentioned above, the baseline uses a plain text editor interface plus dictation. Raw transcripts populate the text content area as a user speaks, after the microphone is activated. The transcript can be edited using iOS's built-in keyboard and cursor interface. To avoid potential confounding effects resulting from differing speech recognition engines, we implemented our baseline interface using the same speech recognition engine used in our prototype (Assembly.ai), as a web interface. For a fair comparison between the baseline and \tool, in our baseline condition we provided ChatGPT through a separate tab in the same web browser---thus ensuring that our results are primarily driven by the affordances of \tool rather than the availability of an LLM. However, we did limit the use of ChatGPT to revising content produced by participants, by discouraging participants from asking ChatGPT to generate lengthy compositions.

\subsection{Participants}
Twelve participants ($N = 12$), seven who self-reported as female and five as male, were recruited from our university campus community through various mailing lists and student organizations. All participants were native-fluent English speakers, and there was varied experience with dictation and LLM use.  P3 and P10 were aged between 25 and 34; all other participants were aged between 18 and 24. No participant had motor impairments that impacted keyboard or dictation usage.
Participant details are presented in Table \ref{Tab:participants}.

\subsection{Experiment Task} 
Participants were asked to \textit{compose and revise a blog post} using the given interface in each condition. We chose this task to encourage participants to compose long-form personal writing and revise it extensively for an audience, in order to address a potential lack of motivation, in a lab study setting, to improve text the participants may not have any reason to feel invested in. Thus participants were asked to choose a topic they were personally interested in, from a provided list of open-ended writing tasks (e.g., ``Think of two dream jobs of yours.'' and ``Think of two academic concepts you find interesting.''). 
For each of the two conditions, participants were asked to create a ``blog post that would be posted on your social media'' about one chosen answer to the topic. 
Note that we had participants use the same topic across conditions (e.g., two dream jobs, not one dream job and one academic concept), to mitigate the potential impact of topic selection on participants' compositions.

\subsection{Procedure}
A lab study session for each participant lasted about 90 minutes, divided into three parts: the pre-study survey, the main study itself, and the post-study interview. The study session was conducted in a quiet room, in person, with an experimenter. Participants accessed \tool via Safari on a researcher-provided 11-inch iPad Pro running iOS 16. There was no hardware keyboard, however, participants were allowed to use the on-screen keyboard as needed.

\paragraph{Pre-study survey.} Participants were first given a brief, initial demographic survey including questions about their familiarity with writing, dictation tools, voice-interface tools, and LLMs.

\paragraph{Main study.}
Participants were not told we made the interfaces. In a counterbalanced order across participants, each participant used both interfaces in two conditions.
For each condition, the participants engaged in 1) a tutorial for training, 2) the task, and 3) a post-task survey. In the training trial, the interviewer spent about 5 to 15 minutes explaining all the functionality of the given condition and provided time for the participant to try and familiarize themselves with each feature. 
Next, to perform the task, the participants were given up to 20 minutes to improve their composition until the time is up, or until they felt satisfied, whichever came first. 
If the participants hit a word count of 800 words or more, or had not done any editing by the 10-minute mark, the experimenter would encourage the participant to begin editing (to ensure enough time for revisions).
In the post-task survey, the participants were asked to assess the usefulness of each feature in the given condition using a five-point Likert scale. They were also asked to estimate what percentage of their time was spent in micro-revision (specifically keyboard-editing), compared with macro-revision. Finally, participants were asked to assess the quality of their text composition on a five-point Likert scale.

\begin{table}[t]
    \small
    \centering
    \caption{Demography and dictation + LLM usage of study participants.} 
    \begin{tabular}{ | p{0.05\columnwidth}| p{0.15\columnwidth} | p{0.16\columnwidth} | p{0.25\columnwidth} | p{0.16\columnwidth}| }
        \hline
        \textbf{ID} & \textbf{Degree} & \textbf{Domain} & \textbf{Dictation Usage} & \textbf{LLM Usage}  \\
        \hline
        P1  & Ph.D. & Comp. Sci. & Yearly & Weekly \\
        P2  & Undergrad & Comp. Sci. & Never & Everyday \\
        P3  & Ph.D. & Comp. Sci. & Never & Never \\
        P4  & Undergrad & Comp. Sci. & Monthly & Weekly \\
        P5  & Undergrad & Business & Weekly & Monthly  \\
        P6  & Undergrad & Cog. Sci. & Never & Weekly \\
        P7  & Ph.D. & Elec. Eng. & Never & Yearly \\
        P8  & Undergrad & Data. Sci. & Never & Monthly  \\
        P9  & Undergrad & English & Monthly & Monthly  \\
        P10  & Ph.D. & Physics & Yearly & Monthly  \\
        P11  & Undergrad & English & Yearly & Never  \\
        P12  & Undergrad & Comp. Sci. & Never & Never  \\
        \hline
    \end{tabular}
    \label{Tab:participants}
\end{table}

\paragraph{Post-study interview.} Following the two conditions,  participants were interviewed regarding their overall experience, for around 25 minutes. Specifically, they were asked about their workflows for each condition and their experiences with the features in both conditions.
Finally, participants were asked for their overall preference between the baseline and the \tool condition.
Each participant was compensated \$30 USD for their participation.

\subsection{Data Collection and Analysis}

We captured three types of data for analysis: a screen recording of the iPad running the \tool application; audio recording of participant and interviewer speech during each study session; and application logs of user interactions with the interfaces. We performed descriptive quantitative analysis of the ratings and logs, and summarized our interview findings in qualitative analysis. 




Qualitatively, we engaged in exploratory data analysis, transcribing all audio recordings, and coding the screen-recording videos to observe participants' workflows and strategies. 
Two of the authors then compared these approaches across participants and categorized their usage through a thematic analysis, relying on a modified form of affinity diagrams~\cite{moggridge2006designing} and service blueprints~\cite{bitner2008service} to document the specific workflows and strategies each participant engaged in and the experiences they communicated with the interviewer.

Additionally, we evaluated participant output text for fluency and grammaticality using the GRUEN automated benchmark~\cite{zhu2020gruen}, and compared our participants' final text with similar ``blog post'' text sampled from a popular corpus of weblog posts~\cite{schler2006effects} to match participant demographics on age and gender.

\section{Results}







This section synthesizes the findings from our quantitative and qualitative analysis. We analyzed the overall workflow based on screen recordings and identified a broad diversity of ways \tool was used. The questionnaires and interviews compared the effectiveness of \tool in supporting writing with speech, and collected feedback on user acceptance. The usage of each feature in \tool was reported in reference to system logs, ratings, and interviews.

\subsection{Text Output Quality}

Across both subjective and objective assessments, 
text output quality did not differ significantly between \tool and the baseline. 

\begin{figure}
    \centering
\includegraphics[width=\columnwidth] {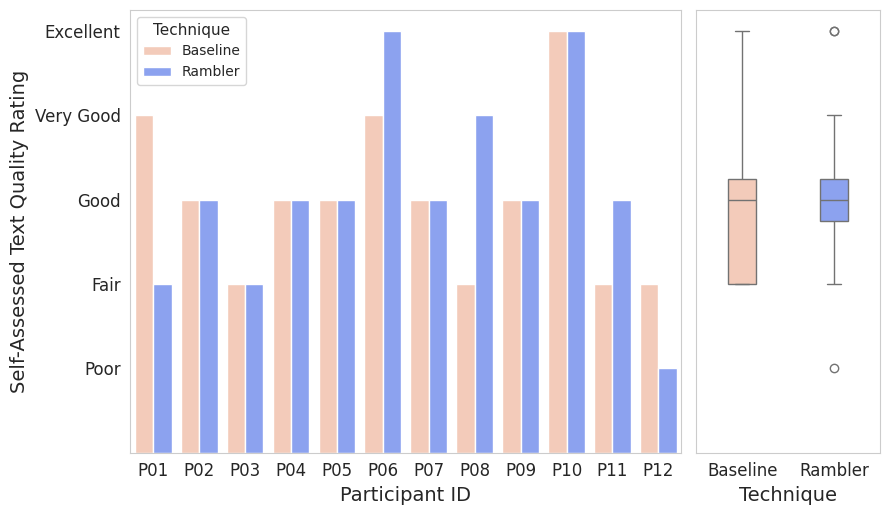}
    \caption{\edits{Self-assessed text quality Likert rating by condition (technique)---Baseline and \tool. Left: individual ratings per participant; Right: aggregated box plot per condition. }}
    \label{fig:text_quality}
\end{figure}

\paragraph{Subjective assessment} We asked participants to rate the quality of the text they submitted as the final outcome for each task. As \edits{shown} in Figure \ref{fig:text_quality}, more than half of the participants rated the outcome quality the same across both conditions. Two participants rated higher for the baseline and three rated higher for \tool. An independent t-test of the Likert ratings revealed no significant difference in perceived text quality across conditions, consistent with the overview we get from the boxplot in Figure \ref{fig:text_quality}. 

\newedit{
\paragraph{Objective assessment} We also performed an objective text quality assessment using an established computational method, GRUEN~\cite{zhu2020gruen}, which calculates a score of fluency based on grammaticality, non-redundancy, and focus. 
We compared the GRUEN scores of the participants' text output using a paired t-test across the two conditions, and found no statistical difference ($p=0.46$, $t(11)=0.77$) between \tool ($M=0.75$, $SD=0.16$) and the baseline ($M=0.70$, $SD=0.26$). 
This is consistent with the subjective assessment, further validating that the text output quality of \tool is comparable with the baseline.} 

\newedit{Additionally, we used GRUEN scores to further assess the text output quality of writing with speech versus traditional writing. We compared our participants' final text output quality across both conditions with an author-demographic-matched sample of 74 Internet blog post texts~\cite{schler2006effects}. Here, we found that our participants' GRUEN scores were significantly higher than the blog posts. Two paired t-tests were performed separately comparing the GRUEN scores of \tool and the blog posts ($p<0.0001$, $t(84)=4.25$), and comparing the baseline and the blog posts ($p=0.0008$, $t(84)=3.48$). 
The GRUEN scores were 83\% 
higher using \tool ($M=0.75$, $SD=0.16$), and 
71\% 
higher using our baseline ($M=0.70$, $SD=0.26$), than in the blog posts sample ($M=0.41$, $SD=0.26$). 
This shows, in terms of fluency, non-redundancy and focus, that LLM-assisted writing with speech in both methods can outperform traditional writing. 
}

\begin{figure*}
  \centering
  \begin{subfigure}{0.45\linewidth}
    \centering
    \includegraphics[width=\linewidth]{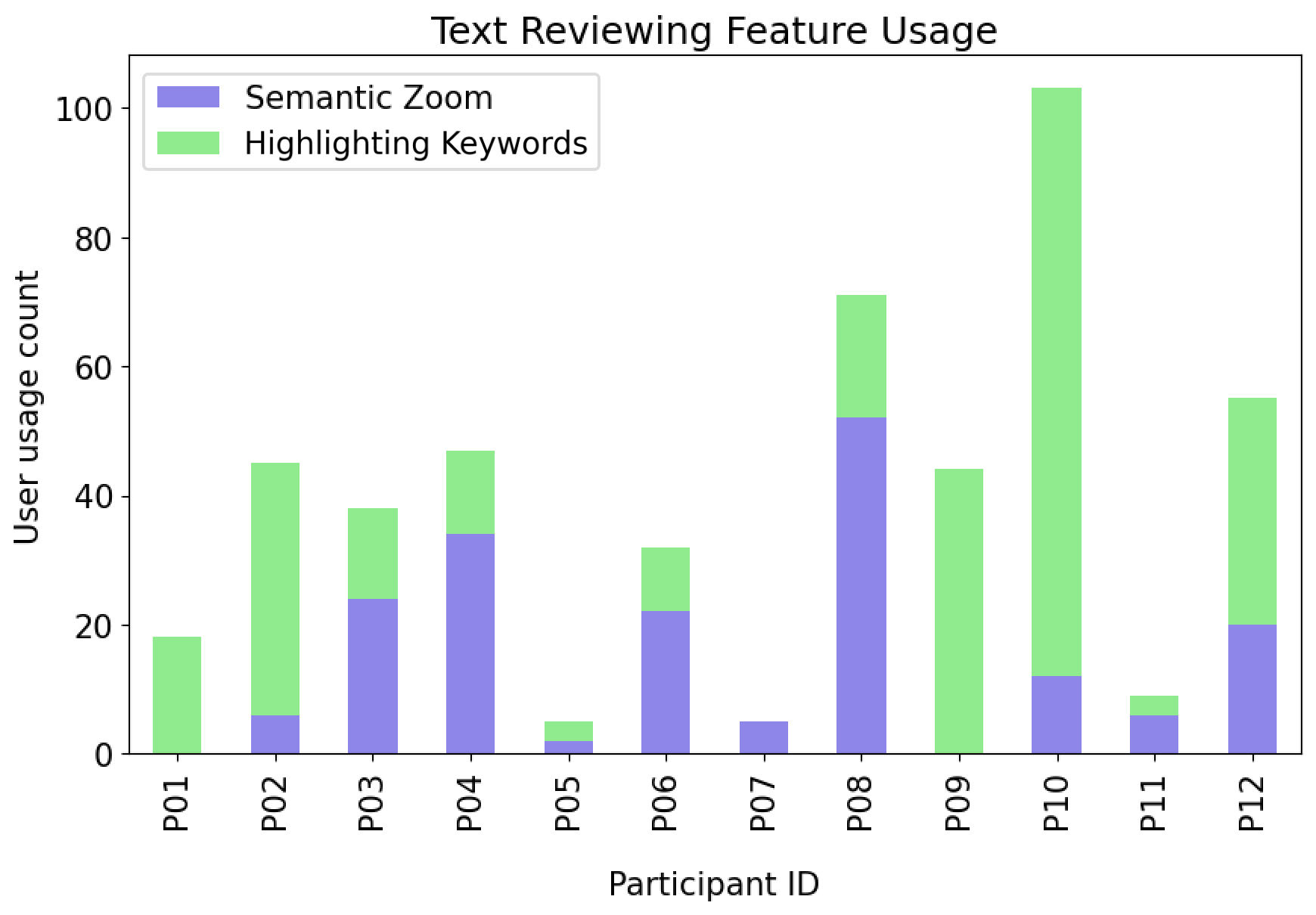}
    \caption{}
    \label{fig:text-reviewing-logs}
  \end{subfigure}%
  \hfill
  \begin{subfigure}{0.45\linewidth}
    \centering
    \includegraphics[width=\linewidth]{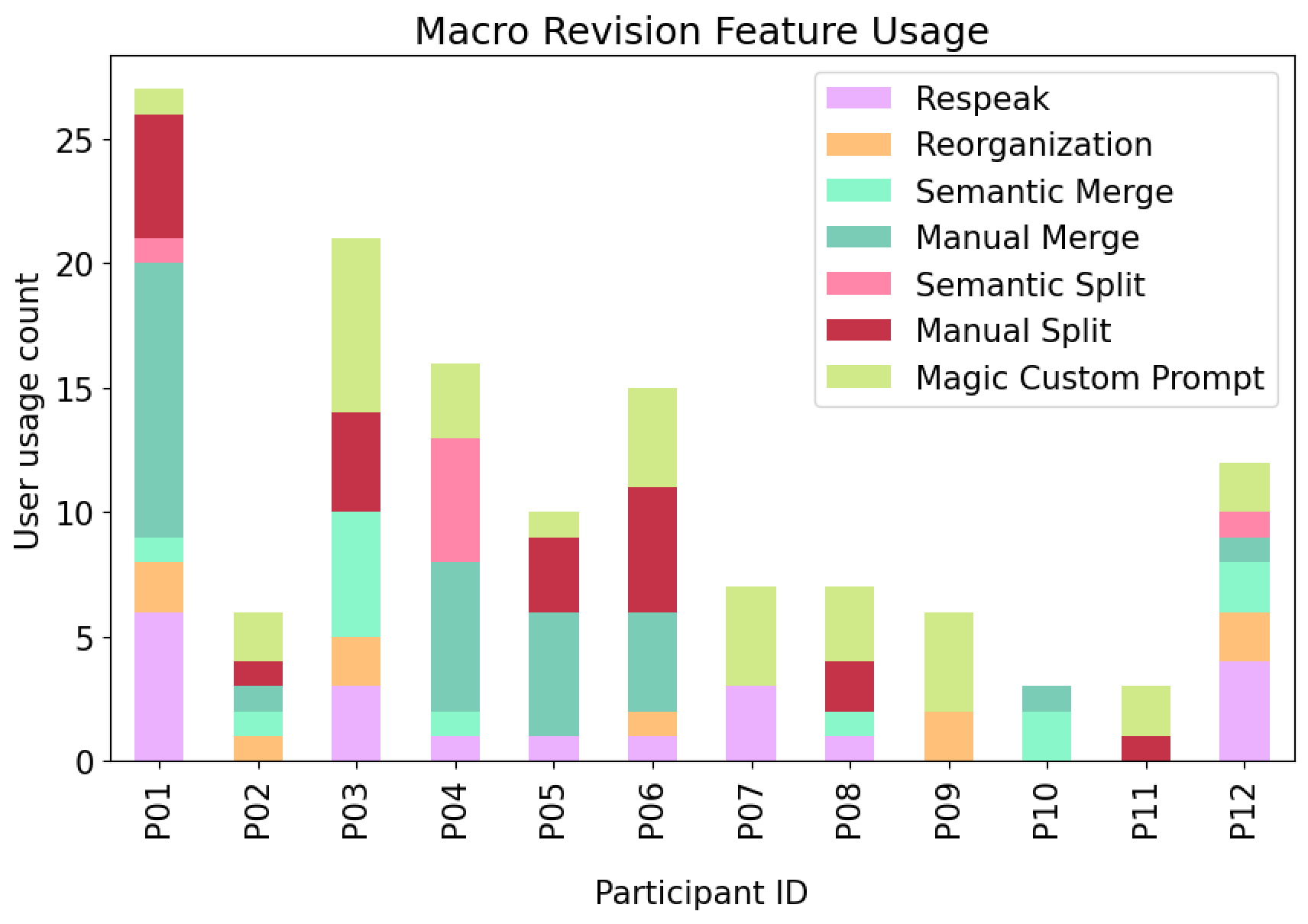}
    \caption{}
    \label{fig:magic-revision-logs}
  \end{subfigure}%
  \caption{ \edits{Stacked bar graph of feature usage counts associated with (a) text reviewing and (b) macro-revision. In (a), one usage count for Semantic Zoom represents \edits{a single} transition from a Semantic Zoom level to an adjacent level (e.g., 50\% summary to 25\% summary). 
  All other usage counts in (a) and (b) represent a single tap, drag, etc. on their respective interfaces.}}
  \label{fig:log}
\end{figure*}

\subsection{Analysis of Workflows and Strategies} 
Participants exhibited a large diversity of approaches in using the various features offered in \tool; no single consistent workflow emerged as the most common. This section summarizes and highlights our findings about participants' workflows and strategies, \newedit{which came from coding the screen recording videos with the themes of editing goals. We excluded P6 from the workflow analysis due to a technical issue that prevented us from capturing their screen recording.} Appendix~\ref{app_timeline} illustrates detailed operations performed by individual participants for each task based on system logs and describes each participant's workflow based on our observation. 


\subsubsection{Overall Composition and Revision Workflow} \label{sec:strategy}

\newedit{There are two known writing styles identified in the writing literature, \emph{structured} and \emph{freeform}, distinguishing whether authors have a narrative planned ahead of time~\cite{10.1145/3532106.3533506} (structured). In writing with speech, we observed two potentially related styles: \emph{content-first}---dictating most content first---and \emph{intertwined}---revising as they compose. This was also observed in previous work about spoken composition~\cite{10.1145/3555758}. As we can see in Figure~\ref{fig:workflow_timeline}, three participants (P8, P10, and P11) followed a content-first approach by dictating most of their text first, before editing through macro- and micro- revisions. Eight participants (P1, P2, P3, P4, P5, P7, P9, and P12) adopted an intertwined approach and switched between dictation and revision more frequently.}

\subsubsection{\newedit{Strategies for Individual Editing Goals}}

\newedit{Our analysis also sheds some light on how participants achieved individual composition and editing goals, 
including \emph{adding, reviewing, reorganizing, rephrasing,} and \emph{removing content}. }

\newedit{
\paragraph{Adding Content}
All 11 participants used dictation to compose text, through creating a new Ramble, as their main way of adding new content. Seven participants (P1, P3, P4, P5, P7, P8, 12) also utilized Respeaking to add content to an existing Ramble. P1 used Manual Split to separate a specific sentence out of a Ramble so that they could add onto it.}

\newedit{
\paragraph{Reviewing content}
Eight participants used either Semantic Zoom, Highlighting Keywords, or Regenerating Summaries to explicitly help review their composition. Four participants (P4, P8, P10, P12) used Semantic Zoom to help review during their process, while three (P2, P7, P11) used Semantic Zoom to only review at the very end of their composition task. One participant (P12) also used Highlighting Keywords by itself as a form of review, while P2 and P10 used Highlighting Keywords combined with Regenerating Summaries to review. }

\newedit{
\paragraph{Reorganizing Content}
As expected, participants made use of the Manual (onscreen keyboard-based) and Semantic (LLM-based) Split and Merge functions to segment content. But we also observed interesting mix-and-match strategies combining manual and semantic operations. 
Three participants (P1, P4, P12) used Semantic Split first, before using Manual Merge on some of the resulting Rambles to achieve a partially Semantic, partially Manual Split. P3 used Semantic Merge and then Manual Split to control some of the uncertainty of what they would get from Semantic Merge. P1 also paired Manual Split and Manual Merge to more precisely resegment their Ramble (by moving the last sentence of the first Ramble into the second Ramble). More uniquely, P10 got rid of all their segmentation through a combination of Manual Merge and Semantic Merge, and P4 experimented on a Ramble through Manual Merge, then Semantic Split, and then used Semantic Merge on the resulting pieces. }

\newedit{
Furthermore, three participants (P1, P3, P4) utilized Magic Custom Prompt to attempt to split their composition: P3 (“make into bulleted list”) and P4 (“Reformat like a design proposal”) did so at the beginning of the task, while P1 (“Can you… organize it into paragraphs”) did so at the end of their task. As Magic Custom Prompt is designed to only modify a single Ramble and Rambles do not store paragraph breaks, these prompts would just create a block of text within a Ramble, and participants would go on to either replace through Respeak Ramble (in the case of P3) or segment it through Semantic Split.} 

\newedit{
Seven out of 11 participants (P1, P2, P3, P4, P9, P12) rearranged their Rambles through drag and drop Reordering, with P1 and P4  also altering the order of text by using Manual Merge on non-consecutive Rambles. Two participants (P1, P3) also more specifically rearranged content within their Ramble: P1 did so through Manual Split and then either reordered or did a non-consecutive Manual Merge, while P3 asked Magic Custom Prompt (on a single Ramble) to ``reflow this text to make sense''.}

\newedit{
\paragraph{Rephrasing Content}
All participants used keyboard editing to make minor changes like fixing typos or rewording phrases. Ten out of 11 participants utilized Magic Custom Prompt to do a more broad rephrasing. Three participants (P2 P8, P12) only sought specific style changes like ``make it more formal'' or ``like a tumblr post''. 
Five participants (P1, P5, P7, P9, P11) looked for a more general revision of the text through less specific requests like ``revise text,'' ``clean the text,'' or ``make this writing more readable'', while two participants (P3, P4) did both. 
Furthermore, five participants tried to achieve a more uniform style across their entire composition: three participants (P2, P3, P8) used similar prompts across all their Rambles, while two participants (P1, P5) used Manual Merge to combine all their Rambles into a single Ramble so that they could apply Magic Custom Prompt onto their entire text in one go. 
Two participants (P3, P12) also used Respeaking to replace entire Rambles.}

\newedit{
\paragraph{Removing Content}
Ten out of 11 participants deleted content from their composition. Seven participants (P1, P2, P3, P4, P5, P8, P12) utilized the Delete Ramble functionality, and three of those participants (P3, P5, P8) specifically combined Manual Split with Delete Ramble to delete words and sentences from the beginning or end of a Ramble. Three participants (P1, P2, P11) utilized keyboard editing to delete whole sentences from their Rambles.}

\subsection{Usage of \tool Functions}
\begin{figure}
    \centering
\includegraphics[width=\columnwidth] {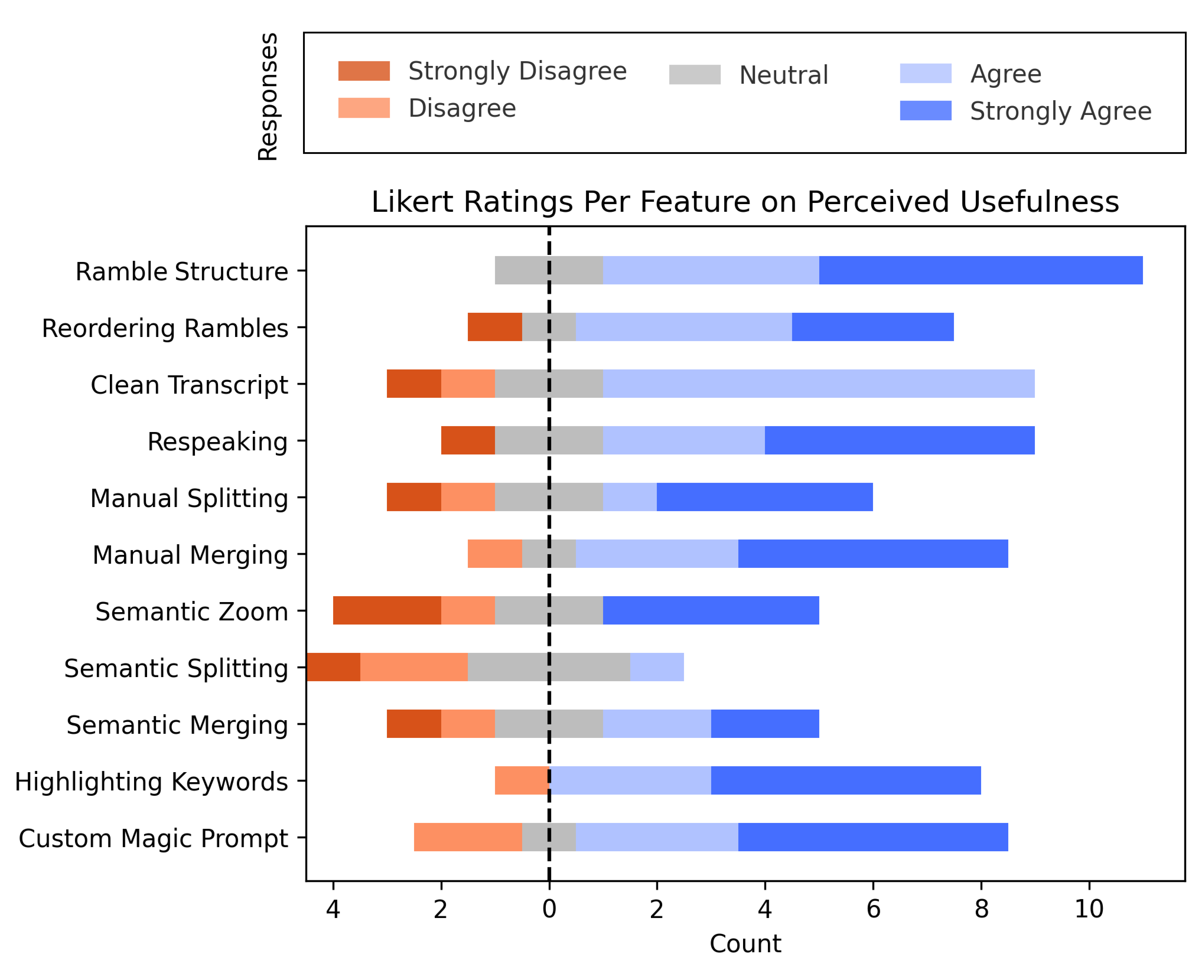}
    \caption{Usefulness Likert Rating per Feature, collected with the statement ``I found $<$feature\_name$>$ useful.'' 
    The total number of ratings per feature varies because some participants indicated that they did not use a particular feature.}
    \label{fig:likert}
\end{figure}

Here, we analyze participants' feature usages from both the system logs and interview data. 
Overall, the participants in the study found \tool to be a useful tool for writing. As detailed in Section~\ref{sec:strategy}, participants invented diverse and creative strategies when using \tool. These diverse strategies led to equally diverse usage of functionalities. 
Figure~\ref{fig:log} visualizes the usage count of each function logged in our system, per participant. As we can see from Figure \ref{fig:text-reviewing-logs}, Semantic Zoom and Highlighting Keywords were the most frequently used features. Overall, the use of features varied largely across participants.
Figure~\ref{fig:likert} shows the subjective ratings of each function in the post-task survey. The following analysis assesses the use of each feature by referring to these ratings and summarizing relevant insights revealed from the interview.  

\subsubsection{The Ramble Structure}
The Ramble structure is the most well-received feature. Participants made good use of the Ramble structure to organize and discretize their writing. While they had varying opinions on how it impacted their writing process, all 12 participants found it useful for its organizational benefits. 
Some participants found the structure more efficient when they had ideas prepared beforehand, while others preferred it for freely brainstorming. P3 and P5 also appreciated that the sections of Rambles mimicked their usual writing style of outlining first, which helped them get used to the structure. 
Furthermore, eight participants mentioned that Rambles helped them to focus and iterate on specific sections of their work. P12 noted that the structure allowed them to separate their ideas and feel more in control of their organization, while P9 found that it enabled them to work in a more methodical manner. Overall, participants found that Rambles helped them to better discretize their composition, which made it easier to visualize and revise.

\subsubsection{Manual Operation Usage for Macro Revision}

Participants found the manual operations helpful, while individual feature usage differs based on their workflow. 

\paragraph{Respeaking}
Seven participants used Respeaking to add new text to a Ramble. Two of those participants used Respeaking to replace entire Rambles. 
Half of the participants found this feature to be a nice addition, and more than half of the participants rated Respeaking to be useful. 
While some mentioned that they could simulate it by deleting and restarting the Ramble, one participant noted that Respeaking is quicker and more convenient.

\paragraph{Reordering Rambles}
Participants' usage of Reordering Rambles depended on their individual workflows. Five participants did not use the Reordering feature.
However, most participants felt Reordering was a useful feature, which is supported in the Likert ratings. 
Interestingly, a couple of participants felt more comfortable rambling knowing that they could easily adjust their ideas afterwards. As P4 put it, “\emph{I was able to just kind of get my ideas on the screen without necessarily like a linear progression, because I knew I could… move things around [after]… the Rambles… [help let] me… freely brainstorm, but then allowing me to kind of put it all together in a way that made sense.}” While several participants pointed out that this can also be done in a normal editor by copy and paste, P12 noticed in a “mobile device [context]…it was a lot easier… copying and pasting text [in the baseline]... was a little bit more clunky”.

\paragraph{Manual Merge and Split}
Manual Merge (via drag-and-drop) and Manual Split (via an onscreen keyboard's return key) were used by participants depending on their workflows. While many participants found Manual Split and Merge to be helpful and useful as further indicated in the Likert ratings, these two features were not considered essential by all. Some participants used them to combine Rambles, take breaks in between dictation, or section out parts of the text for Respeaking. P3 utilized Manual Split as a way for them to regain control over and “rein in” their composition after using Semantic Merge. 
Some participants also found tension between Manual Split and Merge and Semantic Split and Merge, as they preferred one or the other depending on their workflow.

\subsubsection{Semantic Operation Usage for Macro Revision}
\label{sec:findings-macro-revision}

\tool provides a number of LLM-assisted operations, some of which are more well-received than others. Cleaning transcripts is a valuable feature that the majority of participants found helpful. Semantic Zoom is a versatile feature that can be used for a variety of purposes, but it is not as widely used as part of people's strategies. 
Semantic Merge and Manual Split are less commonly used, and some participants stopped using them after a few attempts.

\paragraph{Cleaning Transcripts}
Most participants found the live cleaning of the raw transcript to be helpful and useful. P4 and P12 specifically mentioned how much they appreciated the feature after using the baseline (where they didn't have the automatically cleaned transcript). Participants P11 and P12 were also pleased with how the clean transcript preserved their voice better than what they could get using ChatGPT in the baseline. The accuracy of the clean transcript also made it easier for participants to immediately and actively review, understand, and iterate on their composition. As P11 expressed, the cleaned transcript was so high quality that they were willing to do keyboard edits to it (whereas in baseline they just handed it off to ChatGPT).

\paragraph{Semantic Zoom}
The usage of Semantic Zoom varied depending on each participant's strategy. Most participants used or explored using this feature according to Fig.~\ref{fig:text-reviewing-logs}. It was used for a variety of purposes, such as gauging whether they could query the LLM for more succinctness (P4), review what they talked about (P7, P8, P11), checking if they missed any main ideas (P7), \edits{using generated summaries as part of the final text (P8),} or simply out of curiosity \edits{(P10)}.  
\edits{The relatively high count of usage in Fig.~\ref{fig:text-reviewing-logs} is partially due to the way logs were generated -- if a participant scrolls through the slider it gets registered at each level change.}
\edits{Semantic Zoom was observed to be less used when participants did more keyboard editing, perhaps because participants can gain a good understanding of the text while micro-editing and thus need less assistance for reviewing. 
P3 found Semantic Zoom less helpful because their composition was short enough to not need summaries and felt that the shortest level was too summarized to be useful. On the other hand,} P7 was very impressed with it, and P8, \edits{who used Semantic Zoom as a tool for editing instead of reviewing,} expressed that it was the most useful tool of them all. P8 used the 50\% level as the final submission because it felt like a ``more clean or more postable version'' of their work, and mentioned how they ``liked having more information to go off when I’m reviewing my work.'' \edits{They} also appreciated how easy it was to regenerate the summaries by customizing keywords. 

\paragraph{Semantic Merge and Split}
The usage of Semantic (LLM-assisted) Merge and Split varied depending on participants' workflows. Seven out of 12 participants used \edits{Semantic Merge and three used Semantic Split.} 
Participants used Semantic Split to partition long Rambles (P4), and participants used Semantic Merge to achieve cohesive combinations (P2, P4) or out of curiosity (P3, P10). P3 mentioned it is good for low-stakes where word choice precision doesn’t matter as much. \edits{When it comes to the reasons why Semantic Split was less used, we observed that participants used Manual Split more often than Semantic Split, because it warrants more certainty. As P3 said, ``I sort of use the Manual Splitting as a way to like rein in the Semantic Merge, which I really liked.'' 
Some (P7, P9) felt the task wasn’t long enough for Semantic Split to be useful. P10 was worried about unexpected results, and thus was reluctant to use it.}
We also learned that some participants stopped using these features after a few attempts due to receiving unexpected results (P1).

\paragraph{Magic Custom Prompt}
Most participants made extensive use of Magic Custom Prompt to revise within Rambles: changing tone, reformatting structure, further cleaning up their text composition, \edits{and} even reducing the size of a Ramble.  
One participant spoke with the intention of constructing an outline, then used Magic Custom Prompt to expand their outline into a ``design document'' (P4), which they then split into multiple Rambles for further revision.
Four participants had positive outlooks on the Magic Custom Prompt, though P2 felt that they were relying on it too much. P11 and P12 also liked the Magic Custom Prompt, but primarily because it preserved their writing voice more than ChatGPT in the baseline did when they used keywords to influence the Magic Custom Prompt outcome. 
In contrast, P8 did not like this feature, as they felt that it was not helpful in getting the style or language that they wanted.

\paragraph{Highlighting Keywords and Regenerating Summaries}
Highlighting Keywords (and Regenerating Summaries) was used in different ways by participants, but those who adopted it agreed or strongly agreed it was useful. Participants used keywords in a variety of ways: to understand their readers (P7), improve their focus and understanding of their own writing (P2, P12), highlight keywords to regenerate summaries (P8, P10), and influence their Magic Custom Prompt (P9, P11). P4 found that highlighting keywords could be extra work, as they could change during editing. P11, on the other hand, found that providing keywords as context to the system helped them retain their writing voice.

\subsection{Effectiveness of \tool Compared to Baseline}

The questionnaire and interview reveal that \tool is an effective tool for supporting text composition and revision via speech, especially for tasks that require organizing one's thoughts, iterating on content, and maintaining one's writing voice. \edits{We explicitly asked participants about \tool with respect to reviewing composition, organizing thoughts, supporting iteration, and granting user control.}

\subsubsection{Support for Reviewing the Composition}
Eight out of 12 participants found \tool to be more helpful for reviewing text composition than the baseline system. Half of these participants attributed this to the Ramble structure, which made it easier to discretize, visualize, and review their text composition. The other half found Semantic Zoom to be useful for grasping the main idea of their Rambles and checking that they included or emphasized all the points they wanted to. Additionally, the automatically generated clean transcript made it easier for participants to review their text.

\subsubsection{Support for Organizing Thoughts}
Eleven out of 12 participants felt that \tool was better for organizing their thought process. Most participants attributed this to the Ramble structure and its inherent support for reorganization. For example, P6 said that the Ramble structure made it ``a lot easier... to keep track of what I was doing.'' P1 explicitly said that \tool was ``clearly better for organizing... I was able to move the ideas around. And that was very helpful.'' P4 also found it ``super useful for dragging my thoughts around and kind of restructuring.''
P3 was the only participant who felt that the baseline was better for organizing their thought process. They rationalized this by the free-form nature of the baseline's text editor. Their preferred workflow depends on a nested structure which \tool doesn't support.

\subsubsection{Support for Iteration}
Seven out of 12 participants felt that they iterated more in \tool. This was due to the Ramble structure, the support for macro-revisions, and the automatically generated clean transcript.
The Ramble structure made it easier to discretize text composition and allowed for more flexibility, which made iteration faster. P3 said that \tool was "a lot more ...inviting of ... merging and smashing things together. So I could definitely iterate a lot more."
The clean transcript made it easier to review on the spot and enabled iteration on the go. P10 and P11 explicitly mentioned that they iterated more in \tool because they did not feel motivated to iterate at all in the baseline.

\subsubsection{User Control}
Ten out of 12 participants felt that \tool gave users more control over the content iteration. While some functionality like Manual Merge, Manual Split, and Reorganization were pointed out as mimicable in the baseline, five participants emphasized how the support within \tool made it faster to perform macro-revisions. Knowing that there was support for iteration made participants feel more in control.
Furthermore, the structure of the Rambles helped localize changes. Four participants mentioned that this made it easier to know what part of the text the intelligent text processing was going to be operating on. P8 said that "when you figure out what [\tool{}] does, it is much easier to really customize your composition... If you want more detail, \tool is worth learning."
Some users also noted how they felt more in control of their writing voice. P11 and P12 were pleased at how the clean transcript more accurately preserved their voice than ChatGPT could. P11 explicitly mentioned how keywords helped preserve their voice when they utilized the LLM through Magic Custom Prompt. Among many other participants, they expressed that \tool made it possible for them to ``see [the composition] better and edit it better''.

\subsection{User Acceptance and Use Cases} 

\subsubsection{On using speech to write}

Dictation is a powerful tool for capturing thoughts for writing, but it can take time to learn how to use it effectively. Our post-study interviews revealed mixed opinions on the experience of writing with speech, indicating a learning curve to overcome.
\edits{Three participants who used dictation a few times a month or year (P9, P10, and P11)} found it to be ``a really good way'' to get started and ``capture [your] loose thoughts before they run away'' \edits{(P9). P4 (a few times a month) felt both tensions: it ``kind of just let your ideas flow'' to use dictation, but there's also ``a little build up in my brain of... what should I say first.'' One participant who reportedly used dictation weekly (P5) described their process as ``I speak before I think'' but ``that's not exactly a good thing'' because their ``words are definitely not clean''--- and that \tool encouraged them to ``ramble a lot more''}. 
Finally, some participants with no experience \edits{(three who had never used it before: P2, P3, and P7)} felt that speaking made them think about what to say first and dictation slowed them down. \edits{P6} also mentioned that they did not like having to manually clean up STT transcripts.
\edits{Overall we did not find a consistent trend between dictation experiences and their opinion about it being positive or negative, perhaps due to our limited sample size.}


\subsubsection{On using \tool for their own writing}
Despite the learning curve of using speech for writing, 10 out of 12 participants preferred \tool over the baseline and envisioned themselves using it in the future.
The participant feedback suggests that \tool is a versatile tool well-suited for a variety of tasks. First, we have text editing tasks: four participants mentioned situations where it would be useful to edit or break down text using \tool. One participant said that Semantic Zoom would be useful for digesting content more quickly. Another participant said, ``What makes \tool like better for certain use cases is being able to have like this higher level control of ideas. So I really enjoyed using those macro tools… I feel like this process of just like moving that you feel like you can actually like move an idea… in an abstract sense. I really enjoyed that part of it'' (P4).
Then, we have spoken-language tasks: Five participants mentioned using \tool for transcribing conversations, interviews, or lectures, or composing speeches, presentations, or scripts.
Five participants mentioned using \tool for brainstorming, describing it as a way to ``capture the essence of things''.
Three participants mentioned using \tool for blogging or journaling (e.g., low-stake/shorter tasks).

\section{Discussion}

\subsection{Summary of Findings}

We found that writing with speech can be better supported by \tool compared to the state-of-the-art practice, which is an STT text editor with the help of ChatGPT. 
\tool is shown to be an effective tool for supporting text composition and revision via speech. Specifically, it showed advantages in helping users review spoken compositions and organize their thoughts, encouraging and supporting iteration while enhancing user control. Furthermore, participants used \tool in highly creative and diverse ways, suggesting that \tool is a versatile tool that can be adapted to different writing styles and workflows. 
The most frequently used features were Highlighting Keywords and Semantic Zoom. Macro revision features were mostly well received. Some LLM-assisted operations like Semantic Merge and Semantic Split were used less frequently but creatively. Participants who did use them found them helpful. The Magic Custom Prompt served as a lubricating custom feature used to achieve a variety of editing goals in both rephrasing and reorganizing content. Finally, writing with speech was considered to be a good way to get started and capture loose thoughts. Despite the learning curve for getting used to writing with speech, almost all participants in the study preferred \tool over the baseline and envisioned themselves using it in the future.

\subsection{Effects of Gist-based Interface Features} 

The concept of \emph{gist-based interface} was suggested for spoken text in previous theoretical work \cite{10.1145/3571884.3597134} and explored in a few existing interfaces \cite{Dang22,jiang2023graphologue} although not in this exact term. In this work, we further frame this concept in the context of dictation interfaces for long-form writing. It is operationalized with a few features, including: \emph{gist extraction} through Semantic Zoom and Highlighting Keywords; \emph{gist manipulation} through macro revision operations including Ramble respeaking, Ramble reorganization, Ramble merging, splitting, and custom transformation. 

Our data showed that gist extraction features were the most frequently used. Keywords appeared to be highly versatile: visual grounding, users' self-annotation, customizing summaries, and influencing custom LLM transformation to align content with users' voice. Semantic Zoom \edits{had mixed ratings and was actively employed by fewer participants despite being tried by most. When employed, it was} used in multiple ways including supporting review, serving as a checklist of ideas, and generating concise versions of participants' writing to be included in their final draft. 

Macro revision in essence refers to interaction with text in larger chunks, in particular to support the \emph{conceptual level of text manipulation}, as opposed to micro revision which focuses on verbatim editing word-by-word. In theory, it should serve well for speech-to-text input, considering the speed of speech input, weaker verbatim memory, good contextual correction of ASR, and the difficulty of performing small corrections on mobile. In our study, manual operations for macro revisions were well-used, including Respeaking and Manual Splitting / Merging. Although LLM-assisted Splitting and Merging were less used, participants came up with creative strategies to appropriate these features, showing promising new ways of interaction with text via gist manipulation. Uncertainty about the outcome of these features remains an obstacle and affects users' trust. 

Overall, \tool sets a starting point for exploring the design space of gist-based interfaces for text review and manipulation. Our findings suggest that such interfaces can support useful GUI manipulations without always requiring high input precision.

\subsection{\newedit{Design Implications}}

\newedit{Our findings from the design and evaluation of \tool suggest a number of implications for the future design of systems for writing with speech.
First, our findings reveal that the greatest benefit of using \tool is the affordance of the Ramble structure and macro revisions. Being able to manipulate text in chunks serves mobile platforms well, given that precise input is a challenge in that environment. Spoken text in particular benefits from support for iteration, such as automatic cleaning of disfluency and both LLM-assisted and manual support for reorganizing chunks of text. } 

\newedit{Second, the popularity and high rating of \tool's Magic Custom Prompt reveals a need for customized requests in AI-assisted word processing. As also found in other research efforts embedding LLM capabilities in graphical user interfaces~\cite{10.1145/3586183.3606800,10.1145/3586183.3606772,10.1145/3586183.3606725}, \tool is well-served by having a mixture of deterministic shortcuts (buttons) and custom prompts, offering users control on multiple levels. Users can rely on common operations without worrying about prompt quality, while still being able to take full advantage of AI support via natural language requests. Some participants used Magic Custom Prompt to split or reformat one large Ramble into multiple paragraphs or other structures, indicating the potential usefulness of custom prompting for scopes larger than a single Ramble. However, part of the benefit of having operations applied to individual Rambles is that changes are localized: word processing through a conversational UI (as in our baseline condition) makes it difficult to apply changes at specific text locations and observe changes. }

\newedit{Third, both macro revision and micro revision are critical components of \tool; they go hand-in-hand, and are complementary in serving users' goals. In our user study, keyboard editing was still used by most participants in the \tool condition, even if visibly less than in the baseline condition (see Figure~\ref{app_timeline}).}

\newedit{These implications together also suggest a future of writing interfaces that combine input modalities---systems that can draw on the affordances of speech, keyboard, and other input modes as appropriate, relying on LLM underpinnings to help mitigate impedance mismatches (e.g., by cleaning up disfluencies) and create new affordances (e.g., Semantic Zoom).}

\subsection{Challenges Embedding LLM API Calls in a Latency-Sensitive GUI}

Under the hood, many of \tool's system architecture decisions were driven by the latency capabilities and limitations of the underlying LLM (GPT-4) and transcription (AssemblyAI) APIs. We found GPT-4's latency to be roughly linear to output length; for example, producing a 60-word summary took about 7 seconds.

Given that this would impact user experience in features like Semantic Zoom, we opted to pre-generate all summaries in parallel (and cache them for later) as soon as dictation ended, rather than waiting until users sought them out. We used the \textit{streaming} form of the GPT-4 API to receive summaries. This approach showed users at least \textit{some} summary text right away at each level, even if it would take several seconds to fully complete, improving interactivity.

As an illustration of how latency limitations and LLM capabilities interact to steer interfaces and interactions, consider the following anecdote from our development process: We initially attempted a mechanism for summary levels to update \textit{in real-time} as the user was still dictating text, by sending that dictated-text-in-progress through our GPT-4 summary pipeline. However, this pipeline yielded a new summary with nearly zero word-level consistency with the old summary, even if the difference between new and old transcript was minor. A summary that completely changes every few seconds is not useful, but we found a way to specifically prompt GPT-4 to \textit{build on} the previous existing summary based on the \textit{changes} to the transcribed text from one summary requested to the next. Though this approach now generated useful ``append-style'' summaries, the prompts grew three times as fast as the transcript did---rapidly reaching latencies above 20 seconds, a delay so long as to render them essentially useless.

\section{Limitations}
\label{sec:limitations}

One limitation of this work is that the study was conducted in a lab with an assigned task to participants. This artificial setting does not strongly motivate participants to feel invested in the task outcome. Given the nature of \tool, this poses a challenge as participants are more likely to claim completion rather than continue editing, limiting the data we can collect. Although we designed the task specifically to \edits{try to} mitigate this, there are likely still differences between what we could find in a 90-minute long session where participants were given a topic to write, in comparison with a more realistic task with greater benefit for the participant, in which the participant would feel more invested in the outcome. \edits{Additionally, a relatively small sample size leads to challenges in concluding some of the potential correlations, e.g. potential effects of participants' dictation experiences.}  

\edits{Further, our choice of study device was an iPad tablet, representing one of the mobile platforms suitable for long-form writing. We believe our general concept of supporting writing with speech via rambles, gist extraction, and macro revision can be applied to other mobile devices such as smartphones. However, a smaller screen would demand some UI adjustments. Not all our findings could directly generalize to a smartphone platform. For instance, the feature usages may differ as a smaller screen renders more challenges in content reviewing and manual editing, which points to a potentially greater need for support. Our current system already offers a responsive web interface, we plan to make UI adjustments and run future studies on other mobile devices in particular smartphones.} 

\newedit{In addition, we used a computational method for objective assessment of text quality produced by participants. This method is limited to evaluating fluency, redundancy and focus, therefore does not encompass complete matrices of text quality. It serves as an additional result to triangulate with the subjective assessment of text quality.} 

\edits{Lastly, \tool itself, as a research prototype, also comes with certain limitations. \tool semantic features relying on the GPT-4 API experience some unavoidable latency in the order of seconds. Our reliance on large language models for operations like Semantic Split and Semantic Merge might increase complications about proper attribution of authorship. \tool also does not yet support multiple files or syncing across devices.}


\section{Future Work}
\label{sec:futurework}

Promising directions for future work include a longitudinal, diary study, which would provide participants with \tool for longer and in a more realistic setting \edits{with their own mobile devices}, \edits{potentially yielding insights into use less skewed by novelty effects}. \edits{This diary study should also include a more diverse participant pool, addressing the limitation of our study being focused on an academic context. 

While \tool showed a decent degree of versatility by surfacing highly diverse participant strategies in our lab study, future work shall investigate how to further support these diverse strategies, perhaps by making the features modular and user-customizable by choosing their own combinations. 
 }
Future development of the approach described here could, for example, explore how to support personalization of LLM-powered GUIs through user-defined components for users' own LLM operations. Lastly, here we merely scratched the surface of the design space of possible Gist-based Interfaces, \edits{supporting only a subset of the possible semantic operations made possible by LLMs.} Extending this work to other use cases \edits{and other form factors} beyond writing with speech on mobile devices could, in combination with this work, shed substantial light on new ways of writing more broadly.

\section{Conclusion}

In this paper, we describe the design, implementation, and evaluation of \tool, a versatile gist-based user interface that supports users to iteratively transform spontaneous ``rambling'' text into well-structured writing.
With \tool, users can start writing easily and quickly by speaking ``rambling'' ideas into our interface, seeing the content in clean text, grasping its gist with Semantic Zoom and Highlighted Keywords, and iterating on text content via LLM-assisted macro revisions (i.e., respeaking, merging, splitting, and rearranging). Our evaluation shows several major advantages of \tool compared to a baseline speech-to-text editor with ChatGPT, especially in supporting reviewing, manipulating, and revising the spoken text. The supported macro revision interactions are especially suitable for mobile devices as they relax the constant need for high-precision input for editing text. \newedit{Our work also contributes one example, in the context of interacting with spoken text, in support of the notion that injecting LLM capabilities into a GUI tailored for a specific task---and which also supports user-defined prompts---can outperform relying solely on chat-based LLM interfaces like ChatGPT.} 

\begin{acks}
This project was funded in part by the Berkeley Artificial Intelligence Research Lab - Open Research Commons, and by the National Natural Science Foundation of China - Young Scientists Fund (CityU 62202397). We thank the Android Input Research team at Google for discussions and feedback. 
\end{acks}







\balance

\bibliographystyle{ACM-Reference-Format}
\bibliography{./main.bib}

\appendix

\clearpage

\section{Appendix}
\subsection{Participant Workflows}\label{app_timeline}



\begin{figure*}[h!]
    \centering
    \includegraphics[width=0.7\textwidth] {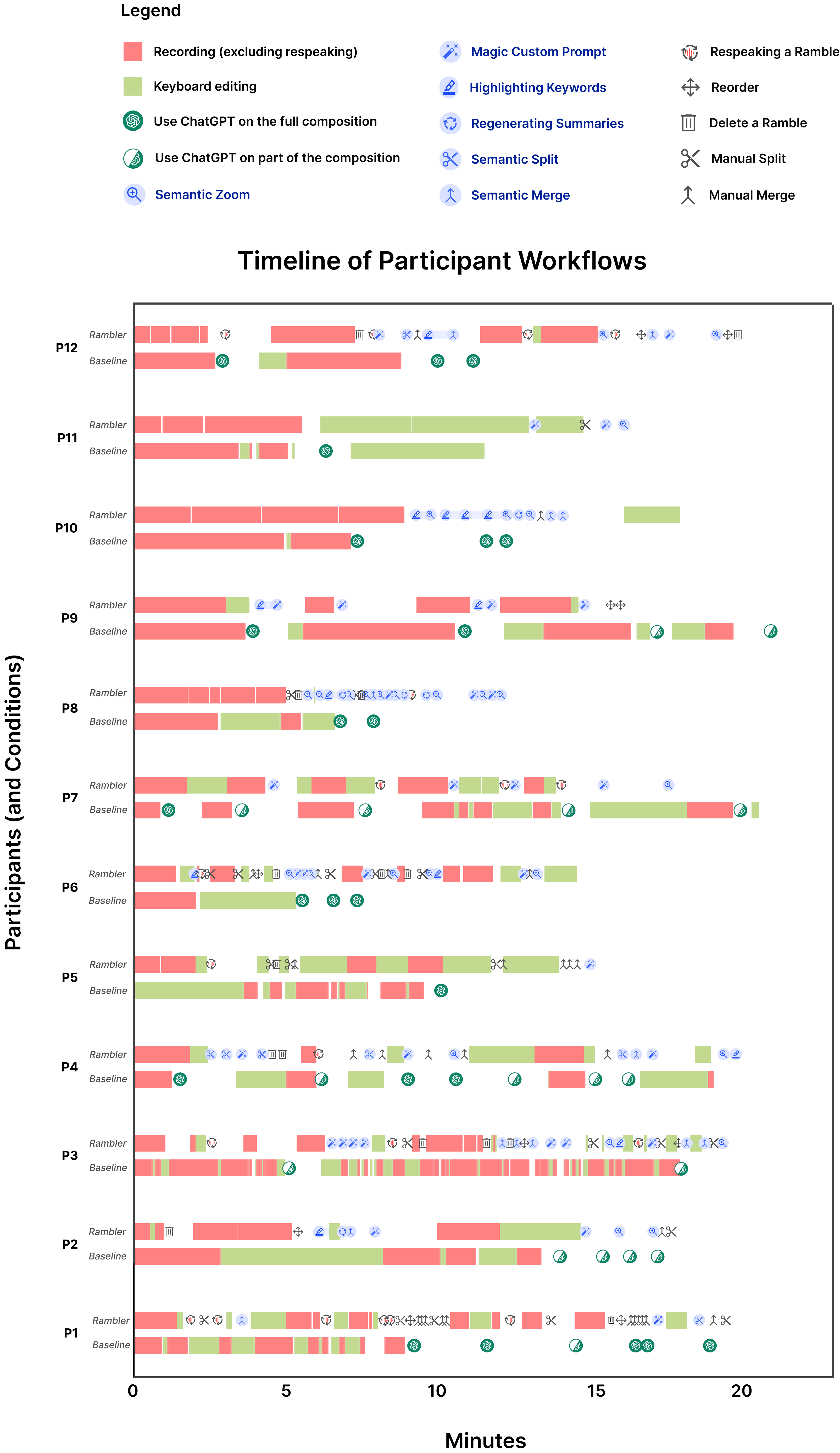}
    \caption{\edits{Timelines visualizing participants' workflow for each task. The icons are manually annotated based on the coding of the screen-recordings of the tasks. The timelines are a qualitative observation of feature usage: consecutive triggers of functions may be annotated as single uses, and obvious accidental triggers for keyword highlighting were omitted.}}
    \label{fig:workflow_timeline}
\end{figure*}

See Figure~\ref{fig:workflow_timeline}, later, for a visualization of participant workflows.


\subsubsection{Detailed Description of Participant Workflows}\label{workflow_list}

\begin{itemize}
    \item \edits{P1 typically started new rambles with a few words, re-speaking to add on, and then keyboard editing it. At the very end, they} manually merged \edits{all rambles together} to apply magic custom prompt, and then they used semantic split to get paragraph breaks.
    \item P2 thought about what to say before actually starting speaking out loud, by pre-dividing their thoughts into Rambles. \edits{They then edited the composition at a higher level through semantic merges and magic custom prompt, before diving into more keyboard editing.}
    \item \deletes{P3 used Rambles like a loose outline, each as a subpoint, then used semantic merging.} \edits{P3 used Rambles like subpoints, creating an outline for themselves. They first recorded a couple of Rambles, and then used a series of magic custom prompts to clean the entire composition, before recording more rambles, and using semantic merge and manual split to restructure the composition.}
    \item P4 first produced a short paragraph and used magic custom prompt to reformat it into a design document (which inherently generated more text). \edits{They then used semantic split and manual merge as they continued to add content through speech. They used more keyboard editing in the latter half of the task.}
    \item P5 did not pre-plan, and spoke as if telling a story; they keyboard-edited each section without much macro revision, but at the end merged everything into a single Ramble to apply the magic custom prompt ``clean up text''.
    \item P6 added \edits{content as it came to mind.} They reorganized, manually merged and split, \edits{applied magic custom prompt to revise tone, and keyboard edited text}. 
    \item P7 spoke in a sequential way\edits{---for each ramble, they would speak, then keyboard edit, and occasionally use magic custom prompt to help revise. At the very end they took a look at the semantic zoom summaries}.
    \item P8 \edits{recorded most of their rambles first,}\deletes{initially revised with keyboard,} then visited 50\% semantic zoom level to see what was there. \edits{They indirectly edited the 50\% zoom level by regenerating summaries, semantically merging, and adding onto rambles until they were happy with the text in the 50\% zoom level summary, which they used for submission.} 
    \item P9 \edits{worked in a fairly linear order.} 
    \edits{After recording each ramble, they would sometimes choose to keyboard edit or alter keywords, and then use magic custom prompt (with keywords as context) to ``clean the text''.} \edits{At the very end, they did some reordering.}
    \item P10 \edits{recorded all their content first, then highlighted keywords in all the rambles and regenerated all the summaries. Then, they semantically merged all of them together to see what this would produce; because the context shifted due to LLM variability, they had to do some keyboard editing at the end.}
    \item P11 \edits{recorded all of their content first,} \edits{and then keyboard edited before using magic custom prompt (using keywords as context). At the very end, they took a look at the semantic zoom summaries.} \deletes{edited keywords and used them as context when asking magic custom prompt to clean up their text, then keyboard edited.}
    \item P12 rambled a lot to branch off ideas and then \edits{restructured using reordering, semantic split and merge, and manual merge.} They highlighted keywords for their own reference. \deletes{also regenerated summaries to be used in their task submission.} 
    
\end{itemize}

\subsection{List of \tool Prompts} \label{app_prompts}

As mentioned in Section~\ref{sec:temp-impl}, we define custom prompts for our LLM-enabled functionality and embody them within our UI elements for ease of use. We list these prompts below.

\subsubsection{Split Text Prompt} 

\begin{quote}
    \textbf{System: } Split the paragraph the user enters into logical, cohesive paragraphs and return the result as a JSON array. Analyze the content and break it up into at least two separate paragraphs (but more where it makes sense). Try to split it into the appropriate number of paragraphs based on the content. Add each paragraph as a separate string element in the JSON array. \\

    Response format: ["Paragraph 1 text", "Paragraph 2 text", "Paragraph 3 text"] \\

    \textbf{User: } [User Ramble text]

\end{quote}

\subsubsection{Merge Text Prompt}

\begin{quote}
    \textbf{System: } You are a paragraph merger bot, capable of merging paragraphs. Please merge the following text into one paragraph of roughly median length as the originals:\\
    
    [Text from selected user Rambles, concatenated with newline]\\
    
    You may use the following keywords to help you merge the text. Ensure that each keyword is in the merged paragraph. \\
    
    The keywords are: [list of keywords]. \\
    
    Again, the resulting paragraph should be roughly the average length of the original paragraphs.
\end{quote}

\subsubsection{Clean Text Prompt}

\begin{quote}
    \textbf{System: } You are a text cleaning bot that clean\edits{s} up the text the user enters by correcting obviously incorrect punctuation and formatting, but otherwise keeping the user's text the exact same. You never ask your own questions. For example, if the user enters:\\
    
    Google. Followed by the neural style transfer method, which became really popular. In a lot of cell phone apps. And these things started to. Show up in art galleries and art exhibitions. And more and more artists start playing with it. And now there's a number of fairly. Um, significant contemporary artists who are also playing, experimented with AI techniques. This work all comes out of the, the academic research literature, and whatnot. So these are images from one academic paper which has really \\
     
    you should return: \\
    
    Google. Followed by the neural style transfer method, which became really popular in a lot of cell phone apps. And these things started to show up in art galleries and art exhibitions and more and more artists start playing with it. And now there's a number of fairly, um, significant contemporary artists who are also playing, experimented with AI techniques. This work all comes out of the, the academic research literature, and whatnot. So these are images from one academic paper which has really\\

    \textbf{User: } [User Ramble text]

\end{quote}

\subsubsection{Summarize Text Prompt}
\label{adx:summarize}

\begin{quote}
    \textbf{System: } You are a professional writer specializing in text summarization. Make a summary of [LEVEL TEXT] of the chunk of the text provided by the user. The summary should reflect the main idea and the most important relationships of the text. You must preserve the same point of view, grammar and tense as the original text. If the text is in the first person, using words like I, you must use the first person as well. If the tone was conversational, you must be human conversational as well. You should use the following keywords to help you determine what to focus the summary on. Ensure that each keyword is in the summary. Try to fit as many as makes sense. Do not include anything else in the response other than the summary. \\
      
    The keywords are: [list of keywords].\\

    \textbf{User: } [User Ramble text]
\end{quote}

where the possible values of [LEVEL TEXT], given that $L=$ the number of words in the user paragraph, are 

\begin{itemize}
    \item 5 words or less 
    \item $L$ / 4 words or less
    \item $L$ / 2 words or less
\end{itemize}

\end{document}